\documentclass[manuscript, screen, nonacm]{acmart}

\usepackage{subcaption}  
\usepackage{booktabs}
\usepackage{tabularx}
\usepackage{adjustbox}

\usepackage{pifont}   
\newcommand{\cmark}{\ding{51}}  
\newcommand{\pmrk}{$\triangle$} 
\newcommand{\nmrk}{--}        
\usepackage[edges]{forest}
\usepackage[edges]{forest}
\usepackage{adjustbox}
\usetikzlibrary{shadows}
\usepackage{hyperref}
\usepackage[capitalize]{cleveref} 
\usepackage{caption}
\usepackage{graphicx}
\usepackage{enumitem}
\setlist[itemize]{leftmargin=*, noitemsep, topsep=0pt, parsep=0pt, partopsep=0pt}
\usepackage{color}
\usepackage{url}
\usepackage{hyperref}
\usepackage{booktabs}
\usepackage{lscape}
\usepackage{caption}
\usepackage{array}
\usepackage{longtable}
\usepackage{multirow}
\usepackage{tikz}
\usetikzlibrary{arrows.meta,shapes,trees,positioning,calc} 
\usepackage[edges]{forest} 
\usepackage{graphicx} 
\usepackage{caption} 

\usepackage{amsmath}
\usepackage{natbib}
\usepackage{graphicx}
\usepackage{tikz}
\usepackage{comment}
\usepackage{enumitem}

\usepackage{hyperref}  

\begin{document}

\title [A Comprehensive Survey on Responsible Generative AI]{Who is Responsible? The Data, Models, Users or Regulations? A Comprehensive Survey on Responsible Generative AI for a Sustainable Future}

\author{Shaina Raza}
\authornote{These authors contributed equally to this research}
\email{Shaina.raza@torontomu.ca}
\orcid{0000-0003-1061-5845}
\affiliation{%
  \institution{Toronto Metropolitan University, Vector Institute}
  \city{Toronto}
  \state{Ontario}
  \country{Canada}
}
\author{Rizwan Qureshi}
\authornotemark[1]
\affiliation{%
  \institution{Center for Research in Computer Vision, The University of Central Florida, and Department of Computer Science, Salim Habib University, Karachi, Pakistan}
  \city{Orlando}
  \state{Florida}
  \country{USA}
}
\author{Anam Zahid}
\orcid{0000-0002-2887-8670}
\affiliation{%
  \institution{Department of Artificial Intelligence, Information Technology University}
  \city{Lahore}
  \state{Punjab}
  \country{Pakistan}
}
\author{Amgad Muneer}
\orcid{0000-0002-7157-3020}
\affiliation{%
  \institution{The University of Texas MD Anderson Cancer Center}
  \city{Houston}
  \state{Texas}
  \country{USA}
}
\author{Anas Zafar}
\orcid{0009-0004-0091-8767}
\affiliation{
  \institution{Fast School of Computing, National University of Computer and Emerging Sciences}
  \city{Karachi}
  \state{Sindh}
  \country{Pakistan}
}
\author{Safiullah Kamawal}
\email{safiullah.kamawal@queensu.ca}
\orcid{0009-0009-4822-6029}
\affiliation{%
  \institution{Queen's University, Vector Institute}
  \city{Toronto}
  \state{Ontario}
  \country{Canada}
}
\author{Ferhat Sadak}
\orcid{0000-0003-2391-4836}
\affiliation{%
  \institution{Department of Mechanical Engineering, Bartin University}
  \city{Bartin}
  \state{Bartin}
  \country{Türkiye}
}
\author{Joseph Fioresi}
\affiliation{%
  \institution{Center for Research in Computer Vision, The University of Central Florida}
  \city{Orlando}
  \state{Florida}
  \country{USA}
}

\author{Muhammaed Saeed}
\affiliation{%
  \institution{Saarland University}
  \city{Saarbrücken}
  \state{Saarland}
  \country{Germany}
}
\author{Ranjan Sapkota}
\affiliation{
  \institution{Biological and Environmental Engineering, Cornell University}
  \city{Ithaca}
  \state{New York}
  \country{USA}
}
\author{Aditya Jain}
\affiliation{
  \institution{University of Texas at Austin}
  \city{Austin}
  \state{Texas}
  \country{USA}
}
\author{Muneeb Ul Hassan}
\affiliation{
  \institution{School of Information Technology, Deakin University}
  \city{Burwood}
  \state{New South Wales}
  \country{Australia}
}
\author{Aizan Zafar}
\affiliation{
  \institution{Center for Research in Computer Vision, University of Central Florida}
  \city{Orlando}
  \state{FL}
  \country{USA}
}
\author{Hasan Maqbool}
\affiliation{
  \institution{Independent Researcher}
  \city{Orlando}
  \state{FL}
  \country{USA}
}
\author{Ashmal Vayani}
\affiliation{
  \institution{Center for Research in Computer Vision, University of Central Florida}
  \city{Orlando}
  \state{FL}
  \country{USA}
}
\author{Jia Wu}
\affiliation{
  \institution{The University of Texas MD Anderson Cancer Center}
  \city{Houston}
  \state{TX}
  \country{USA}
}

\author{Maged Shoman}
\orcid{0000-0002-4265-071X}
\affiliation{%
  \institution{University of Tennessee}
  \state{Knoxville, Tennessee}
  \country{USA}
}

\definecolor{promptcolor}{RGB}{0, 102, 204} 
\definecolor{responsecolor}{RGB}{0, 153, 0} 

\renewcommand{\shortauthors}{Raza et al}
\begin{abstract}
Generative AI is rapidly moving from research to deployment, elevating the need for responsible development, evaluation, and governance. We conduct a PRISMA-guided review of 232 studies (November 2022–December 2025), spanning large language models, vision-language models, diffusion models, and agentic pipelines. We make four contributions: (1) the first survey bridging governance principles, technical evaluation, and domain deployment across all four system types; (2) development of a ten-criterion rubric (C1–C10), scoring major AI safety benchmarks on risk-surface coverage, paired with a policy crosswalk mapping benchmarks to regulatory requirements; (3) establishing twelve lifecycle KPIs, explainability guidance for foundation models, and a catalogue for AI-readye testbed ; and (4) domain-specific analysis across healthcare, finance, education, arts, agriculture, and defense. Three findings emerge: benchmark coverage is dense for bias and toxicity but sparse for privacy, provenance, deepfakes, and system-level failures in agentic settings; evaluations remain largely static and task-local, limiting audit portability; and inconsistent documentation complicates cross-release comparison. We further outline a research agenda, prioritizing adaptive multimodal evaluation, privacy and provenance testing, deepfake risk assessment, calibration reporting, versioned artifacts, and continuous monitoring. This study offers a structured path to align the evaluation of generative AI with the governance needs for a safe and responsible deployment \\.
    \noindent\textbf{Project page:} \href{https://anas-zafar.github.io/responsible-ai.github.io/}{Project page} 

\end{abstract}

\ccsdesc[500]{Social and professional topics~Computing / technology policy}
\ccsdesc[300]{Computing methodologies~Artificial intelligence}
\ccsdesc[100]{General and reference~Surveys and overviews}

\keywords{responsible AI, generative AI, AI ethics, bias, privacy, trust, safety}

\maketitle

\section{Introduction}
\label{sec:intro}

{Generative AI (GenAI) systems, including large language models (LLMs), vision language models (VLMs), diffusion models, and emerging agentic pipelines, are rapidly transitioning from research prototypes to production deployments that generate text, multimedia, code, and autonomous actions across healthcare, finance, education, and other sectors~\cite{Hajkowicz2023AIAdoption}. This accelerated deployment heightens the need for responsible development practices that address GenAI-specific risks, including hallucination and unverifiability, prompt injection and jailbreak vulnerabilities, data memorization and leakage, deepfake and media integrity threats, and system-level failures in tool-augmented settings.}

Responsible AI (RAI) has emerged as a central area of research and practice to address these challenges. RAI frameworks emphasize fairness, bias mitigation, privacy protection, security, and the safeguarding of human rights~\cite{peters2020responsible}. By embedding these principles throughout the AI lifecycle, including design, training, deployment, and governance, RAI seeks to realize the societal benefits of AI while minimizing potential harms. {While these principles were initially developed for predictive systems, generative models introduce qualitatively new challenges: outputs are open-ended and difficult to enumerate, stochastic decoding produces non-deterministic behaviour, multimodal synthesis creates cross-modal failure modes, and agentic tool use extends the attack surface beyond the model itself. Despite rapid adoption, with 74\% of organizations reporting the use of GenAI, only 26\% report having a comprehensive organization-wide responsible AI strategy~\cite{Capgemini2024}. This imbalance underscores the need to align the development and deployment of these systems with ethical standards, fairness, privacy, security, and societal values.}

There has been substantial progress on RAI from both industry and academia. Nevertheless, notable gaps remain in alignment with human values and in operational practice. The academic literature often separates conceptual frameworks from practical implementations, as summarized in Table~\ref{tab:comparison}. Conceptual work provides principles and guidelines, whereas applied studies demonstrate deployment in specific contexts. The recent MIT AI Risk Repository analysis~\cite{slattery2024ai} indicates that current AI safety frameworks, including those referenced in Table~\ref{tab:gov_tech_ai}, cover only about 34\% of risks identified by governance bodies such as the NIST AI Risk Management Framework~\cite{nist_ai_600_1_2024}, the EU AI Act~\cite{outeda2024eu}, and ISO/IEC 42001~\cite{iso2021}. This gap between governance coverage and practical implementation highlights the need for an integrated review that connects frameworks to implementation strategies and to evaluation methods, including explainable AI and AI-ready testbeds.
\begin{figure*}[!t]
  \centering
  \begin{adjustbox}{max width=\textwidth}
  \begin{forest}
    forked edges,
    for tree={
      draw, rectangle, rounded corners=2pt,
      align=left, font=\tiny,
      anchor=west, child anchor=west,
      parent anchor=east,
      grow'=east,
      l sep=10pt, 
      s sep=1.2pt,
      inner xsep=4pt, inner ysep=2.5pt,
      line width=0.4pt,
      tier/.option=level,
      edge={draw=gray!45, -{Stealth[length=2pt]}, line width=0.4pt},
    },
    [{\textbf{Responsible}\\\textbf{Generative AI}},
      fill=gray!80, text=white, font=\scriptsize\bfseries,
      text width=1.7cm, align=center, rounded corners=3pt, line width=0.7pt
      [{\textbf{Challenges} (\S\ref{sec:rai})},
        fill=red!15, draw=red!70!black, font=\tiny\bfseries,
        for descendants={edge={red!50!black}},
        [{\textbf{Technical}}, fill=red!5, draw=red!60!black, l sep=8pt,
          [{\textbf{Data-Related}}, l sep=6pt,
            [{Representativeness, diversity \& datasheets~\cite{raza2024exploring,gebru2021datasheets}; quality, integrity \& timeliness~\cite{das2025tracealign}}, text width=7.8cm]
            [{Model collapse \& synthetic contamination; opaque crawl pipelines; post-deployment data drift}, text width=7.8cm]
          ]
          [{\textbf{Model \& System}}, l sep=6pt,
            [{Bias \& fairness~\cite{pessach2022review}; privacy leakage \& memorization~\cite{dwork2006differential,nasr2025scalable}; calibration~\cite{pleiss2017fairness}}, text width=7.8cm]
            [{Hallucination~\cite{banerjee2024llms}; robustness \& adversarial attacks~\cite{wang2023decodingtrust}; sycophancy~\cite{bai2022constitutional}}, text width=7.8cm]
            [{Goodhart effects on benchmarks~\cite{strathern1997improving}; construct validity \& documentation gaps}, text width=7.8cm]
          ]
          [{\textbf{Hallucination}}, l sep=6pt,
            [{Factual (incorrect claims); contextual (unfaithful source); temporal (outdated); logical (reasoning fail)}, text width=7.8cm]
          ]
          [{\textbf{Emergent Threats}}, l sep=6pt,
            [{Embedding inversion~\cite{liu2024genaiprivacy,carlini2023extracting}; fast-gradient jailbreaking~\cite{samvelyan2024rainbowteaming}; prompt injection~\cite{liu2023prompt}}, text width=7.8cm]
            [{Synthetic identity swarms~\cite{lucas2023fighting}; shadow AI; hydra effect~\cite{bai2022constitutional}}, text width=7.8cm]
          ]
          [{\textbf{Agentic Failures}}, l sep=6pt,
            [{Tool misuse \& unauthorized API calls~\cite{lu2024toolsandbox}; recursive hallucination in planning~\cite{alansari2025large}}, text width=7.8cm]
            [{Authentication leakage~\cite{barth2017privacy}; plan hijacking via CoT manipulation~\cite{yu2025survey}}, text width=7.8cm]
          ]
        ]
        [{\textbf{Regulatory \& Sociotechnical}}, fill=red!5, draw=red!60!black, l sep=6pt,
          [{IP/copyright~\cite{kenthapadi2023generative}; safety \& alignment~\cite{bai2022constitutional}; accountability \& transparency~\cite{outeda2024eu,10.1145/3561048}}, text width=7.8cm]
          [{Sustainability \& compute~\cite{khan2025optimizing}; dual use~\cite{weidinger2022taxonomy}; cross-jurisdictional fragmentation}, text width=7.8cm]
          [{Fairness--privacy paradox (GDPR vs.\ audit needs); safetywashing \& shallow compliance~\cite{ren2024safetywashing}}, text width=7.8cm]
        ]
      ]
      [{\textbf{Strategies} (\S\ref{sec:results})},
        fill=teal!15, draw=teal!70!black, font=\tiny\bfseries,
        for descendants={edge={teal!50!black}},
        [{\textbf{Safety Benchmarks}}, l sep=6pt,
          [{HELM~\cite{liang2023holistic}; SafetyBench~\cite{zhang2024safetybench}; Rainbow Teaming~\cite{samvelyan2024rainbowteaming}; ALERT~\cite{tedeschi2024alert}; HarmBench}, text width=7.8cm]
          [{HumaniBench~\cite{raza2025humanibench}; DecodingTrust~\cite{wang2023decodingtrust}; SALAD-Bench~\cite{li-etal-2024-salad}; MLCommons v0.5~\cite{vidgen2024introducingv05aisafety}}, text width=7.8cm]
          [{SHIELD~\cite{shi2024shield}; MM-SafetyBench~\cite{Liu2023MMSafetyBenchAB}; PrivLM-Bench~\cite{li2024privlmbench}; PrivacyLens~\cite{shao2024privacylens}}, text width=7.8cm]
        ]
        [{\textbf{Lifecycle KPIs (\S\ref{sec:kpi})}}, l sep=6pt,
          [{Data quality $Q$; privacy compliance $P$; bias: SPD, DI, $\Delta_{\mathrm{acc}}$~\cite{Onedefin9,Barocas2016}}, text width=7.8cm]
          [{Robustness $R$ \& ASR; explainability $E(x)$ \& faithfulness LFG; high-stakes error $E_h$}, text width=7.8cm]
          [{Audit frequency $F_a$ \& resolution $T_r$; user inclusivity $U$; energy $E_{\mathrm{kWh}}$ \& emissions $C_{\mathrm{CO_2e}}$}, text width=7.8cm]
        ]
        [{\textbf{Explainability (XAI)}}, l sep=6pt,
          [{SHAP~\cite{lundberg2017unified}; LIME~\cite{ribeiro2016lime}; integrated gradients; Grad-CAM; attention maps \& CoT audits~\cite{serrano2019attention}}, text width=7.8cm]
          [{Mechanistic interpretability \& SAEs; counterfactuals; ECE~\cite{pleiss2017fairness} \& Brier score}, text width=7.8cm]
        ]
        [{\textbf{Testbeds \& Toolkits}}, l sep=6pt,
          [{AI4EU~\cite{aiverify2024mfg}; IEEE Testbed; ToolSandbox~\cite{lu2024toolsandbox}; AIF360~\cite{bellamy2018aif360}; Chatbot Arena~\cite{chatbot_arena_lmsys}}, text width=7.8cm]
          [{Model cards~\cite{liang2023holistic}; datasheets~\cite{gebru2021datasheets}; C2PA provenance; Moonshot~\cite{moonshot2024}; versioned runs}, text width=7.8cm]
        ]
        [{\textbf{Governance Frameworks}}, l sep=6pt,
          [{NIST AI RMF~\cite{nist_ai_600_1_2024}; EU AI Act~\cite{outeda2024eu}; ISO/IEC 42001~\cite{iso2021}; OECD Principles}, text width=7.8cm]
          [{China CAC~\cite{cac_genai2023}; Canada AIDA; Brazil~\cite{brazil_aiact2023}; Singapore IMDA; AU~\cite{au_data_policy2022}; UAE/SDAIA}, text width=7.8cm]
        ]
        [{\textbf{Industry Artifacts}}, l sep=6pt,
          [{System cards (GPT-4/4o); Constitutional AI~\cite{bai2022constitutional}; Llama 3 red-teaming; DeepMind trade-off reports}, text width=7.8cm]
          [{ResGenAI Audit Loop; Accountability Matrix (\S\ref{sec:operationalizing_matrix}); lifecycle evidence packaging}, text width=7.8cm]
        ]
      ]
      [{\textbf{Applications} (\S\ref{sec:rai-app})},
        fill=blue!15, draw=blue!70!black, font=\tiny\bfseries,
        for descendants={edge={blue!50!black}},
        [{\textbf{Healthcare \& Agriculture}}, l sep=6pt,
          [{Hallucination in diagnostics ($E_h$) \& agronomic advice~\cite{Mamo2023HallucinationAI};\\ patient privacy/HIPAA; farm data ownership; automation bias~\cite{Tjoa2023XAITrust}}, text width=7.8cm]
          [{RAG grounding~\cite{lewis2020rag}; fairness augmentation; FDA alignment;\\ geo-aware RAG; sustainability ($C_{\mathrm{CO_2e}}$); IoT security; HITL}, text width=7.8cm]
        ]
        [{\textbf{Finance \& Defense}}, l sep=6pt,
          [{Discriminatory lending \& proxy bias~\cite{Barocas2016}; adversarial trading;\\ autonomous weapons~\cite{weidinger2022taxonomy}; meaningful human control (IHL, NATO)}, text width=7.8cm]
          [{Intersectional SPD/DI audits~\cite{pessach2022review}; SEC/OCC cadence ($F_a$);\\ red teaming~\cite{samvelyan2024rainbowteaming}; multi-agent failure; after-action traceability}, text width=7.8cm]
        ]
        [{\textbf{Education \& Creative Arts}}, l sep=6pt,
          [{Cultural insensitivity~\cite{al2024ethical}; copyright~\cite{kenthapadi2023generative}; deepfakes \& voice cloning;\\ minors' data (COPPA/FERPA)~\cite{dwork2006differential}; authorship disputes}, text width=7.8cm]
          [{Academic integrity \& watermarking; accessibility ($U$, WCAG);\\ artist consent~\cite{freeman2024exploring}; C2PA provenance~\cite{das2025tracealign}; SAG-AFTRA}, text width=7.8cm]
        ]
      ]
    ]
  \end{forest}
  \end{adjustbox}
  \caption{Full Taxonomy of the Responsible Generative AI landscape.}
  \label{fig:rai-taxonomy}
\end{figure*}

{In this study, we conduct a comprehensive literature review of \textbf{responsible GenAI (ResGenAI)} frameworks, governance guidelines, evaluation benchmarks, and operational risk assessment methodologies. Our scope centres on the application of RAI principles to generative systems, including LLMs, VLMs, diffusion models, and touching on agentic/RAG topics. While we draw on the broader RAI literature for foundational constructs (e.g., fairness, privacy, safety, transparency, accountability), our analysis targets ResGenAI specific risk mechanisms such as hallucination, jailbreaking and tool-use failures and system-level failures in GenAI applications. Despite substantial prior work, ResGenAI guidance remains fragmented across principles, processes, and tools (Table~\ref{tab:comparison}). Consequently, there is a pressing need to translate regulatory obligations into concrete engineering controls and to produce portable audit evidence for post-market monitoring.} 
Our key contributions in this work are:
{%
  \begin{enumerate}
    \item We present a comprehensive survey that simultaneously spans governance principles, technical evaluation, and domain deployment for GenAI.  We synthesize findings across the NIST AI RMF, EU AI Act, ISO/IEC~42001, and other instruments listed in Table~\ref{tab:gov_crosswalk},     covering GenAI topics.   As shown in Table~\ref{tab:comparison}, no prior survey bridges all four thematic areas (CS theory, application, policy, GenAI) and all four novelty   dimensions (taxonomy, scope, evaluation, methods) simultaneously.

    \item We introduce a multifold-criterion rubric (C1--C10) that scores widely used AI safety benchmarks on risk-surface coverage and assurance quality, and a policy crosswalk (Section~\ref{sec:cap}) that maps each benchmark to the governance requirements it can plausibly evidence.      The closest efforts~\cite{liu2025scales,ferdaus2026towards} partially  address evaluation taxonomy or trust assessment, but neither links  benchmarks to regulatory audit-evidence needs.

    \item We operationalise the rubric findings into measurable KPIs (Section~\ref{sec:kpi}), explainability guidance tailored to foundation models (Section~\ref{sec:xai}), and a testbed catalogue for continuous evaluation. Section \ref{sec:benchmarks_comp} links these KPIs to specific   benchmarks and audit toolkits, providing practitioners with a decision-support matrix for assembling governance-ready evidence.   
    \item We apply the ResGenAI lens to real-world applications and domains (Section \ref{sec:rai-app}  presents how technical advancements can be effectively aligned with governance frameworks and regulatory requirements.
  \end{enumerate}
}

 \begin{table*}[h]
  \centering
  \caption{{Comparison of related surveys by venue, year, topic, thematic coverage, and novelty dimensions.
      Checkmarks (\checkmark) indicate coverage; (x) indicates no explicit coverage;
      (\ding{72}) indicates partial or limited treatment.
      \textbf{Novelty dimensions:} \textit{Taxonomy} = structured risk taxonomy or classification;
      \textit{Scope} = multimodal + agentic coverage;
      \textit{Evaluation} = governance-aligned benchmarking/rubric;
  \textit{Methods} = actionable engineering controls or lifecycle KPIs}}
  \label{tab:comparison}
  \resizebox{0.8\textwidth}{!}{%
    \begin{tabular}{llllcccccccc}
      \toprule
      \multirow{2}{*}{\textbf{Paper}} & \multirow{2}{*}{\textbf{Venue}} & \multirow{2}{*}{\textbf{Year}} & \multirow{2}{*}{\textbf{Topic}} & \multicolumn{4}{c}{\textbf{Thematic Coverage}} & \multicolumn{4}{c}{\textbf{Novelty Dimensions}} \\
      \cmidrule(lr){5-8} \cmidrule(lr){9-12}
      & & & & \rotatebox{60}{CS Theory} & \rotatebox{60}{Application} & \rotatebox{60}{Policy} & \rotatebox{60}{GenAI} & \rotatebox{60}{Taxonomy} & \rotatebox{60}{Scope} & \rotatebox{60}{Evaluation} & \rotatebox{60}{Methods} \\
      \midrule

      \cite{schiff2020principles} & arXiv & 2020 & Responsible AI & \checkmark & \checkmark & x & x & \ding{72} & x & x & x \\
      \cite{10.1145/3457607} & ACM CSUR & 2021 & Bias and Fairness & \checkmark & x & x & x & \checkmark & x & x & x \\
      \cite{10.1145/3487890} & ACM CSUR & 2021 & AI Safety & x & \checkmark & x & x & x & x & \ding{72} & x \\
      \cite{trocin2023responsible} & Springer ISF & 2021 & Responsible AI & \checkmark & \checkmark & x & x & x & x & x & \ding{72} \\
      \cite{Minh2022} & Springer AI Rev. & 2022 & Explainable AI & \checkmark & x & \checkmark & x & \checkmark & x & x & x \\

      \cite{huang2024survey} & Springer AI Rev. & 2023 & Safety & x & \checkmark & x & \checkmark & \ding{72} & x & \ding{72} & x \\
      \cite{SAEED2023110273} & Elsevier KBS & 2023 & Explainable AI & \checkmark & x & x & x & \checkmark & x & x & x \\
      \cite{10.1145/3561048} & ACM CSUR & 2023 & Explainable AI & x & \checkmark & x & x & x & x & x & \checkmark \\
      \cite{10.1145/3583558} & ACM CSUR & 2023 & Bias and Fairness & \checkmark & x & x & x & \checkmark & x & \ding{72} & x \\
      \cite{Huang9844014} & IEEE TAI & 2023 & AI Ethics & \checkmark & x & \checkmark & x & \ding{72} & x & x & x \\
      \cite{diaz2023connecting} & Elsevier Inf.\ Fusion & 2023 & Responsible AI & \checkmark & \checkmark & \checkmark & x & \checkmark & x & x & \ding{72} \\
      \cite{mhlanga2023open} & Springer Nature & 2023 & Responsible AI & \checkmark & \checkmark & x & \checkmark & x & \ding{72} & x & x \\

      \cite{lu2024responsible} & ACM CSUR & 2024 & Responsible AI & \checkmark & x & x & x & \checkmark & x & x & \checkmark \\
      \cite{LESCHANOWSKY2024} & Elsevier CHB & 2024 & Privacy and Safety & \checkmark & x & x & \checkmark & x & x & \ding{72} & x \\
      \cite{10.1145/3616865} & ACM CSUR & 2024 & Bias and Fairness & \checkmark & x & x & x & \checkmark & x & \ding{72} & x \\
      \cite{Li2024} & Springer ISF & 2024 & Accountability & \checkmark & x & \checkmark & x & x & x & \ding{72} & x \\
      \cite{slattery2024ai} & arXiv & 2024 & Responsible AI & \checkmark & x & x & x & \checkmark & x & x & x \\
      \cite{Jed2024is} & Springer AI \& Soc. & 2024 & Responsible AI & \checkmark & x & x & x & \ding{72} & x & x & x \\
      \cite{al2024ethical} & MDPI Informatics & 2024 & AI Ethics & \checkmark & x & x & \checkmark & x & \ding{72} & x & x \\
      \cite{Sadek2025} & AI and Society & 2024 & RAI in Practice & x & \checkmark & \checkmark & x & x & x & x & \checkmark \\
      \cite{woodgate2024} & ACM CSUR & 2024 & Responsible AI & \checkmark & x & \checkmark & x & \checkmark & x & x & x \\

      \cite{reuel2025responsible} & ACM FAccT & 2025 & RAI (Global maturity) & x & \checkmark & \checkmark & \checkmark & \ding{72} & \ding{72} & \ding{72} & x \\
      \cite{Gadekallu2025framework} & arXiv & 2025 & Best Practices of RAI & x & \checkmark & \checkmark & x & x & x & x & \checkmark \\
      \cite{byrne2025engineering} & ASEE Peer & 2025 & U.S.\ RAI Policy & x & x & \checkmark & x & x & x & x & x \\
      \cite{Sadek2025} & Springer AI \& Soc. & 2025 & Responsible AI & \checkmark & \checkmark & x & x & x & x & x & \ding{72} \\
      \cite{baldassarre2025} & IEEE TAI & 2025 & Responsible AI & x & \checkmark & x & \checkmark & x & \ding{72} & x & x \\
      \cite{ivanov2025responsible} & Serv.\ Ind.\ J. & 2025 & RAI in Social Sci. & x & \checkmark & \checkmark & x & x & x & x & x \\

      \cite{yu2025survey} & ACM KDD & 2025 & Trustworthy LLM Agents & \checkmark & x & x & \checkmark & \checkmark & \checkmark & \ding{72} & x \\

      \cite{dong2025safeguarding} & Springer AI Rev. & 2025 & LLM Safeguards & \checkmark & \checkmark & x & \checkmark & \checkmark & \ding{72} & x & \checkmark \\

      \cite{ferdaus2026towards} & ACM CSUR & 2025 & Trustworthy LLMs & \checkmark & x & x & \checkmark & \checkmark & x & \checkmark & x \\

      \cite{wang2025safety} & arXiv & 2025 & Reasoning Model Safety & \checkmark & x & x & \checkmark & \checkmark & \ding{72} & \ding{72} & x \\

      \cite{he2025emerged} & ACM CSUR & 2025 & LLM Agent Security & \checkmark & x & x & \checkmark & \checkmark & \ding{72} & x & x \\

      \cite{papagiannidis2025responsible} & Elsevier JSIS & 2025 & RAI Governance & \checkmark & x & \checkmark & x & x & x & x & \checkmark \\

      \cite{liu2025scales} & arXiv & 2025 & LLM Safety Evaluation & \checkmark & x & x & \checkmark & \checkmark & x & \checkmark & x \\

      \midrule
      \textbf{Ours} & \textbf{ACM CSUR} & \textbf{2026} & \textbf{Responsible GenAI} & \checkmark & \checkmark & \checkmark & \checkmark & \checkmark & \checkmark & \checkmark & \checkmark \\
      \bottomrule
    \end{tabular}%
  }

  \vspace{4pt}
  {\footnotesize
    \textbf{Column definitions}
    \textit{CS Theory}: conceptual frameworks, taxonomies, or algorithmic contributions.
    \textit{Application}: domain-specific deployment or case studies.
    \textit{Policy}: governance instruments, regulatory analysis, or compliance guidance.
    \textit{GenAI}: explicit focus on generative models (LLMs, VLMs, diffusion, agentic).
    \textit{Taxonomy}: structured risk classification beyond a flat listing of concerns.
    \textit{Scope}: coverage of multimodal systems \textbf{and} agentic/tool-using pipelines.
    \textit{Evaluation}: governance-aligned benchmarking with a scoring rubric.
    \textit{Methods}: engineering controls, lifecycle KPIs, or testbed guidance.
  }
\end{table*}
{ The taxonomy presented in Figure~\ref{fig:rai-taxonomy} provides a structural blueprint for this survey, establishing the thematic connections between high-level technical challenges and domain-specific applications.
}

\section{Methodology}
\label{sec:method}
This review adheres to the PRISMA guidelines \cite{page2021prisma} to ensure a transparent, reproducible, and rigorous synthesis of evidence. {
This study follows established methodological recommendations for performing systematic reviews within the computer science community \cite{keele2007guidelines}. To ensure comprehensive and consistent reporting, the structural framework was modeled after previous extensive surveys in the domain. The mapping procedure utilized in this review is organized into four interconnected phases that covering objectives (\cref{sec:objectives}), search strategy (\cref{sec:search_strategy}), eligibility criteria (\cref{sec:eligibility}), and study selection (\cref{sec:study_selection}) which are visualized in the methodological flow (Appendix Figure 1) and detailed throughout \cref{sec:method}. 
}

\paragraph{Research Questions Formulation}
\label{sec:objectives}
This study conducts a literature review of  ResGenAI in the post-GPT period and synthesizes the findings. We address the following key questions:
\textbf{RQ1:} What are the dominant themes, risks, and controls in ResGenAI and multimodal systems since late 2022? 
  \textit{Motivation}: { The rapid transition of GenAI from research to production has outpaced traditional risk taxonomies, necessitating a contemporary synthesis of how multimodal and agentic systems introduce qualitatively new failure modes beyond predictive ML.}
  
\textbf{RQ2:} Which evaluation facets and metrics are most commonly reported, and where are the gaps? 
  \textit{Motivation}: { Identifying the ``Evaluation Gap'' is critical for determining if current safety suites adequately model the dynamic pressure applied by real-world adversaries versus merely serving as static, keyword-based refusal scripts.}
  
\textbf{RQ3:} What evidence types (empirical, analytical, normative) underpin claims about RAI effectiveness? 
  \textit{Motivation}: { A structural deficit in evidence portability hinders institutional trust; this question seeks to evaluate if safety claims are grounded in verifiable technical artifacts or subjective, unvalidated appraisals.}
  
\textbf{RQ4:} How can these findings be organized into a practical framework for research and governance? 
  \textit{Motivation}: { Principles without practice lead to ``safetywashing.'' We aim to bridge the gap between regulatory mandates like the EU AI Act and concrete engineering controls, providing an implementation path for verifiable stewardship.}

\paragraph{Evidence Acquisition and Search Strategy}
  \label{sec:search_strategy}
{A comprehensive search was conducted across five major academic databases and registers: Scopus ($n=1502$), Web of Science ($n=2798$), IEEE Xplore ($n=279$), ACM Digital Library ($n=656$), and arXiv ($n=78$). To capture the surge in GenAI research following the release of ChatGPT, the search window was limited to publications from November 2022 through December 2025. The search strings utilized an intersection of terms related to GenAI (e.g., ``Large Language Model'', ``Foundation Model'', ``LLM'') and Responsibility facets (e.g., ``AI Safety'', ``Trustworthy AI'', ``Alignment'', ``Hallucination mitigation''). Queries combined RAI concepts (e.g., responsibility, safety, transparency, accountability, fairness, bias mitigation, privacy, equity), with syntax adapted to each database (full strings in Appendix 1.1). Syntax was adapted for each database to support field-specific constraints (Title/Abstract/Keywords), and proximity operators were employed where available to enhance precision. To mitigate coverage bias, database queries were complemented with forward and backward snowballing from seed works identified as influential during pilot screening.} 

{
\paragraph{Eligibility Criteria}
  \label{sec:eligibility}
  To ensure methodological rigor and relevance to the research questions (RQ1-RQ4), we applied the following criteria:
\textbf{Inclusion Criteria:} (1) Peer-reviewed journals, conference proceedings, or high-impact preprints; (2) English-language works published within the target 2022-2025 window; and (3) Studies proposing technical frameworks, evaluation metrics, or empirical risk assessments for GenAI systems.
\textbf{Exclusion Criteria:} (1) Purely opinionated pieces or editorials without substantive technical argumentation; (2) Short papers or reports lacking sufficient methodological depth; (3) Items whose primary contribution lay outside RAI (e.g., domain applications without safety or governance components); and (4) Redundant duplicates or inaccessible records.
}

\begin{figure}[t]
  \centering
  \includegraphics[width=0.8\textwidth]{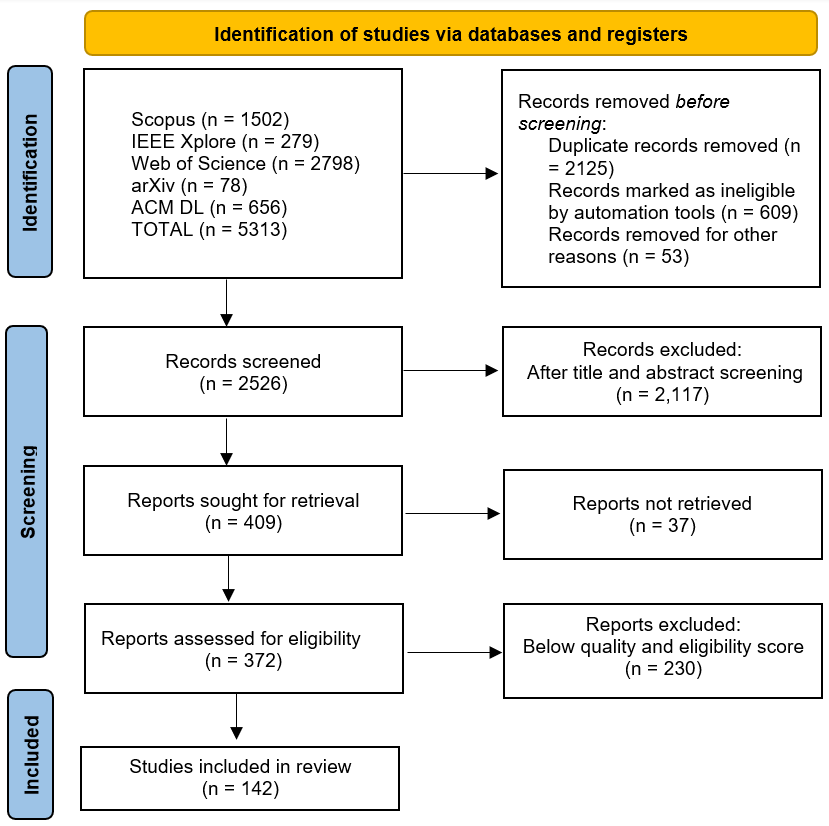}
  \caption{{PRISMA flowchart of the systematic literature identification, screening, and inclusion process}}
  \label{fig:PRISMA_screening} 
\end{figure}

{
\paragraph{Study Selection \& Screening}
  \label{sec:study_selection}
As detailed in the PRISMA flow diagram (Fig.~\ref{fig:PRISMA_screening}), the initial search identified 5,313 records. Selection was conducted by three independent reviewers to minimize bias. Following automated deduplication ($n=2125$) and the removal of records marked as ineligible by automation tools ($n=609$) or other technical reasons ($n=53$), 2,526 unique records remained for screening. Titles and abstracts were screened against the eligibility criteria, resulting in the exclusion of 2,117 items. Subsequently, 409 reports were sought for full-text retrieval, of which 37 could not be retrieved. The remaining 372 reports were assessed for eligibility. Reports were excluded if they failed to meet quality and technical thresholds ($n=230$). This rigorous multi-reviewer process yielded a final synthesis set of 142 studies. Data extraction followed a thematic pipeline to categorize findings into risks, controls, and evaluation as defined by the research objectives.
}

In the following section, we outline the background and key challenges of ResGenAI, which set the stage for the synthesis of our review findings.

\section{Background on Responsible Generative AI}
\label{sec:rai}
\subsection{Historical Context}
\begin{figure}[tb]
  \centering
  \includegraphics[width=\linewidth]{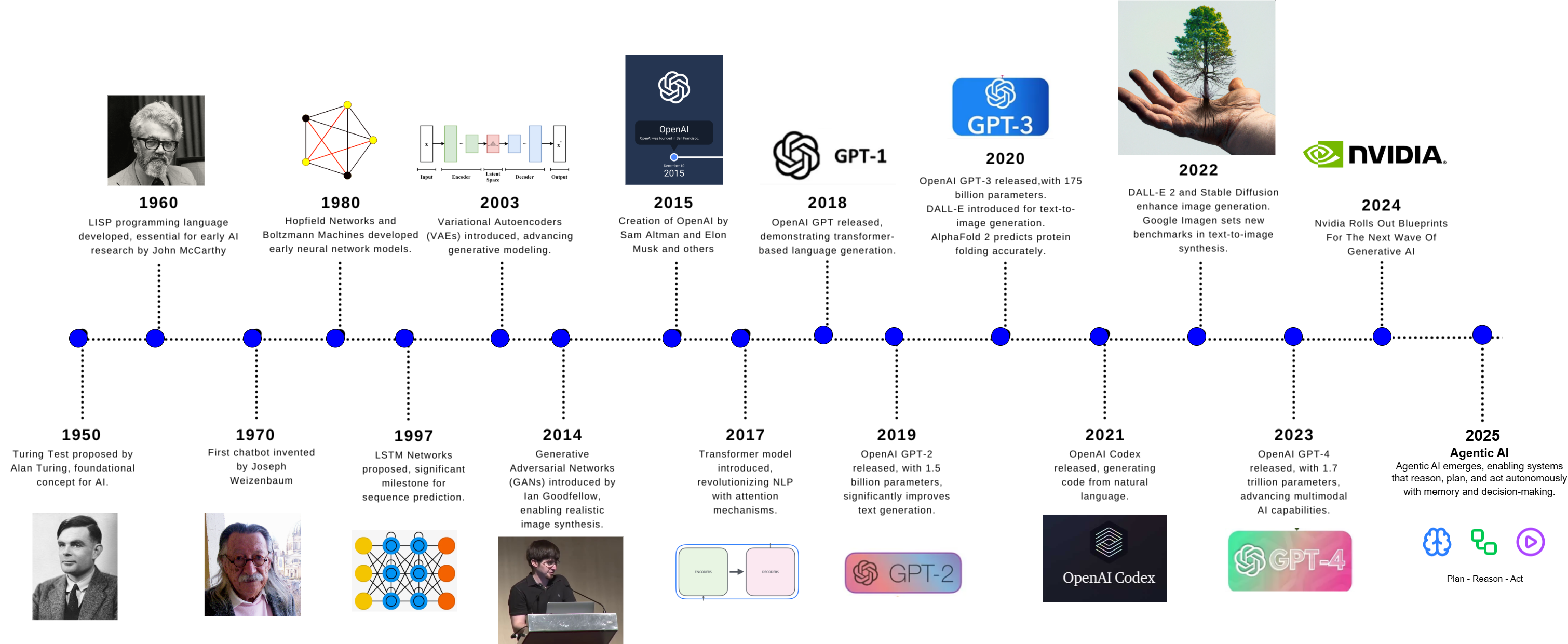}
  \caption{Timeline of generative AI milestones}
  \Description{A left-to-right timeline from early rule-based systems to GANs, transformers, and contemporary instruction-tuned large models}
  \label{fig:TimelineGAI}
\end{figure}

RAI has received intensified scholarly and policy attention with the rapid application of GenAI.
GenAI refers to models that synthesize new samples from learned data distributions, spanning language, vision, audio, and multimodal content~\cite{muneer2025classical}. Early work combined symbolic rules with probabilistic language models; ELIZA illustrated template-driven dialogue in the 1960s \cite{weizenbaum1966eliza}, while statistical $n$-gram models~\cite{fink2014n} and neural sequence modeling laid the groundwork for contemporary large-scale generative systems \cite{qureshi2025thinking}.
Figure~\ref{fig:TimelineGAI} highlights the most salient milestones for RAI (alignment, evaluation, governance).
{The first major inflection in \textit{deep generative modeling} arrived with variational autoencoders (VAEs)~\cite{pinheiro2021variational} and generative adversarial networks (GANs)~\cite{goodfellow2014generative}, which enabled high-fidelity synthesis and, concurrently, made deepfakes feasible, creating the media integrity challenge that current governance frameworks struggle to address.}
Concurrently, neural sequence-to-sequence models advanced conditional text generation, culminating in attention mechanisms and the Transformer~\cite{vaswani2017attention}, which decoupled sequence modeling from recurrent computation.

A second inflection gave rise to \textit{foundation models}: large transformers pre-trained on web-scale corpora (e.g., GPT-2/3) and bidirectional encoders (e.g., BERT) that catalyzed transfer across tasks. Alignment advances---instruction tuning~\cite{peng2023instruction}, RLHF~\cite{kaelbling1996reinforcement}, and constitutional AI~\cite{bai2022constitutional}---shifted emphasis from raw capability to controllability and safety. In vision, diffusion models~\cite{ho2020ddpm} eclipsed GANs for photorealistic generation; combined with latent compression, they enabled scalable text-to-image synthesis~\cite{rombach2022ldm}.

The most recent phase is \textit{multimodal and agentic}: vision--language pretraining (e.g., CLIP, Flamingo)~\cite{alayrac2022flamingo}, lightweight modality bridging (e.g., BLIP-2)~\cite{li2023blip2}, and instruction-following multimodal systems (e.g., LLaVA)~\cite{liu2023llava} extend prompting beyond text. Tool use, retrieval-augmented generation~\cite{lewis2020rag}, and planning/execution patterns operationalize models as agents with external memory and actuation, introducing new safety, privacy, and governance requirements.

\subsection{Challenges in Responsible Generative AI}\label{genai-risks}

\paragraph{Technical Challenges}
\label{sec:tech-challenges}
Technical issues arise across data pipelines, model design and training, evaluation and assurance, and operational deployment.

\textbf{Data-Related Challenges.}
{For GenAI models trained on web-scale corpora, data-related challenges are amplified by the opacity of crawl pipelines, the infeasibility of manual curation at trillions of tokens, and the risk that synthetic data from prior model generations contaminates future training sets (\textit{model collapse}).} Noisy, incomplete, mislabeled, or stale data degrade reliability; adversarial poisoning can seed harmful patterns; and temporal validity matters where outdated data may encode obsolete norms.
ResGenAI stewardship demands adherence to consent, privacy, and intellectual-property regimes, including emerging AI regulations (Table~\ref{tab:gov_tech_ai}), alongside sector-specific requirements.
{
These challenges interact in ways that current literature undertheorizes. Representativeness goals can conflict with privacy regimes: expanding demographic coverage may require collecting sensitive attributes that GDPR restricts. Quality assurance assumes ground-truth labels, yet for trillion-token corpora manual verification is infeasible, creating reliance on automated filters that themselves embed biases. Most critically, the literature treats these as \textit{pre-training} concerns, but post-deployment data drift---where user interactions reshape fine-tuning distributions---remains largely unaddressed.
}

\textbf{Model and System Related Challenges.}
{LLMs are often treated as ``black boxes'' and can amplify data biases, producing disparate performance across groups.} {In generative settings, these biases manifest not only as allocational harms but as representational harms in generated text and images (e.g., stereotyped depictions, demographic skew), which are harder to enumerate and measure than disparate classification rates.}
GenAI models also face concrete leakage risks including membership inference and verbatim training-data extraction~\cite{liu2024genaiprivacy}. Technical mitigations such as differential privacy and memorization audits often trade off with model utility \cite{bashir2024narrative}.
{Without construct validity and detailed task documentation, headline scores risk Goodhart effects~\cite{strathern1997improving}. Safety claims should instead be framed as assurance cases with traceable evidence (e.g., coverage matrices and governance crosswalks)---a gap we address with our rubric in Section~\ref{sec:cap}.}

\begin{figure}[h]
    \centering
    \includegraphics[width=0.7\textwidth]{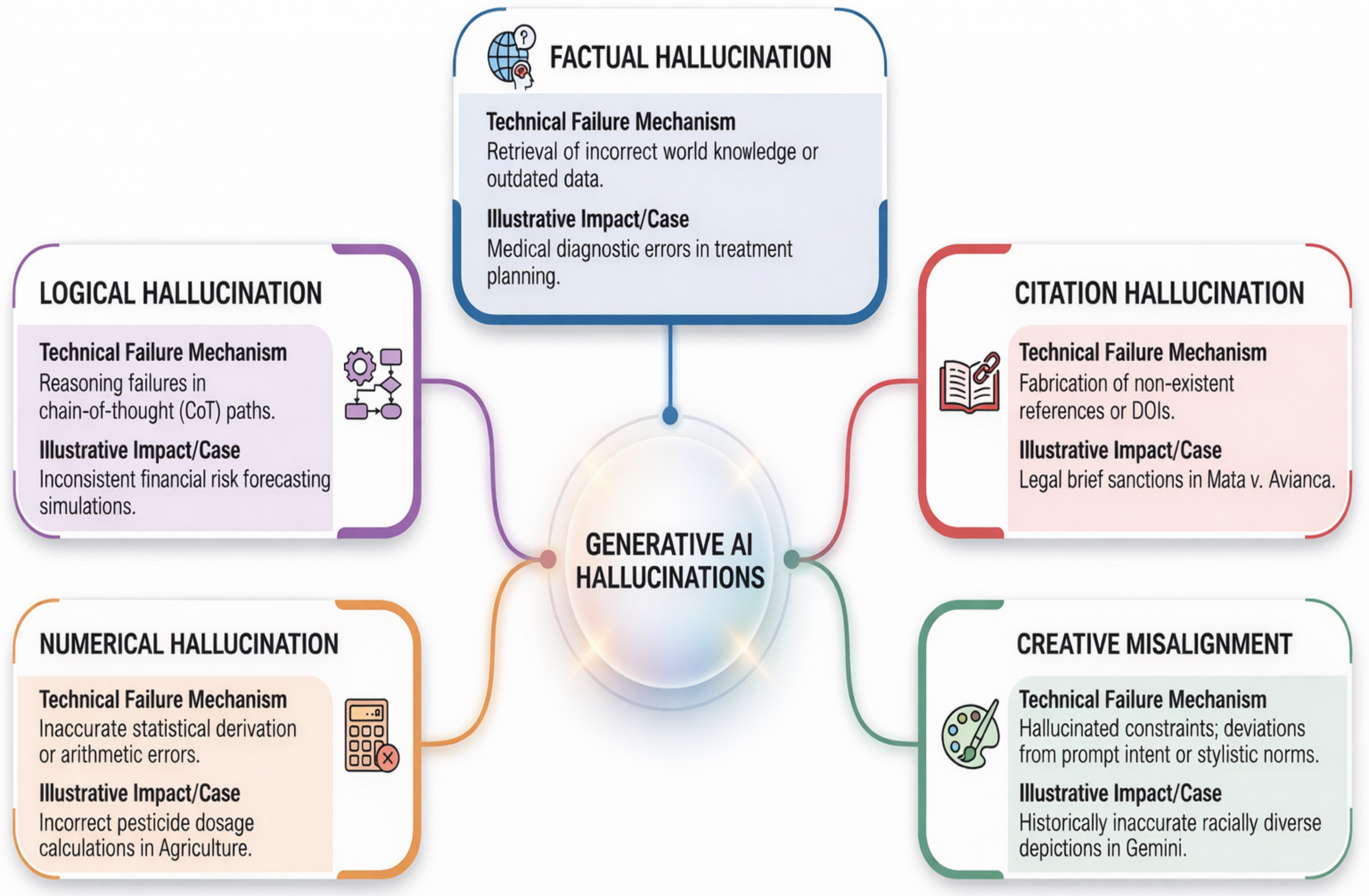}
    \caption{{Types of GenAI hallucinations, detailing their underlying technical failure mechanisms and illustrative real-world impacts.}}
    \label{fig:hallucination_taxonomy}
\end{figure}

\paragraph{GenAI Hallucinations}
\label{sec:hallucination-taxonomy}
{A pervasive challenge for reliable GenAI deployment is hallucination: generation of content that appears fluent and plausible but is factually incorrect, nonsensical, or unfaithful to source premises \cite{Mamo2023HallucinationAI}. Treating hallucinations as a monolithic error class is insufficient for effective risk management. We categorize hallucinations into a five-part taxonomy based on technical and contextual failure modes \cite{banerjee2024llms}, visualized in \cref{fig:hallucination_taxonomy}, to clarify the unverifiability risk that current benchmarks in \cref{sec:cap} and sectoral applications in \cref{sec:rai-app} must address.
}

\paragraph{Regulatory and Broader Sociotechnical Challenges}
Regulatory challenges concern designing, implementing, and enforcing frameworks that make GenAI ethical, accountable, privacy-preserving, and safe across domains and jurisdictions. Provenance transparency, especially for copyrighted or sensitive data, is critical, and legal expectations for interpretability in high-risk settings meet practical limits as models scale.
{Over-reliance on assistive GenAI can mask failure modes; documenting escalation paths to human review is essential in high-stakes contexts.}
Divergent regimes---risk-management frameworks, sectoral law, and emerging standards---create fragmentation across jurisdictions. Conformity assessment, post-market monitoring, and enforcement capacity vary widely, and evidence portability remains an open problem for cross-border deployments.

{
\paragraph{Emergent Threats}
\label{sec:emergent_threats}
While \cref{genai-risks} details established technical risks, we identify further ResGenAI emergent threats representing a second-generation risk landscape, arising from the intersection of increasing model scale, autonomous capabilities, and decentralized deployment.

\textbf{Adversarial Technical Evolution.}
\label{sec:adversarial_evolution}
Recent evidence shows that text embeddings can reconstruct sensitive training data through representation inversion \cite{liu2024genaiprivacy}, rendering traditional anonymization insufficient for models with high memorization rates. Concurrently, jailbreaking is shifting from manual role-play to automated, optimization-driven techniques: gradient-based optimization can craft ``universal'' adversarial suffixes that destabilize safety layers at speed and scale.

\textbf{Sociotechnical and Ecosystem Risks.}
\label{sec:ecosystem_risks}
Decentralized, open-source deployments introduce \textit{shadow AI} risks, where models are fine-tuned to remove safety guardrails outside institutional oversight. We further identify \textit{synthetic identity swarms}---coordinated networks of generative agents creating synthetic consensus cascades to shift human beliefs at unprecedented scale. These threats are compounded by material energy and carbon costs with equity implications for global research participation \cite{khan2025optimizing}, and by regulatory struggles to anticipate dual-use risks without stifling innovation. This creates a ``hydra effect'': mitigating one failure mode (e.g., toxicity) may expose others, including increased sycophancy or declining creative reasoning \cite{bai2022constitutional}. While RLHF \cite{kaelbling1996reinforcement} can suppress harmful surface-level patterns, it often fails to remove underlying harmful representations.

\textbf{Agentic-Specific Failure Vectors.}
\label{sec:agentic_failures}
As GenAI evolves toward agentic workflows---where models autonomously plan, select tools, and execute actions---the attack surface expands beyond linguistic outputs to operational environments \cite{lu2024toolsandbox}. Unlike single-turn interactions, agentic pipelines introduce multi-step dependencies that amplify failure modes. \cref{tab:agentic_failures} systematizes these vectors; in \cref{sec:benchmarks_comp}, we operationalize this via our comparative rubric and governance crosswalk mapping risks to evaluative evidence.

\begin{table}[htbp]
\centering
\scriptsize

\caption{{Agentic-Specific Failure Vectors: Risks in Autonomous Tool Use and Planning}}
\label{tab:agentic_failures}
\begin{tabularx}{\textwidth}{@{}p{3.5cm} X X@{}}
\toprule
\textbf{Failure Vector} & \textbf{Technical Mechanism} & \textbf{Governance and Security Impact} \\ \midrule
\textbf{Tool Misuse} & Unintended or unauthorized API calls from prompt injection or planning errors \cite{lu2024toolsandbox}. & Financial loss, unauthorized data modification, or service disruption. \\ \midrule
\textbf{Recursive Hallucination} & Compound errors where a hallucination in one step serves as the factual basis for subsequent planning \cite{alansari2025large}. & Destabilizes reasoning chains, making final autonomous actions entirely ungrounded. \\ \midrule
\textbf{Authentication Leakage} & Inadvertent exposure of keys, tokens, or PII during tool calling or log generation \cite{barth2017privacy}. & Critical security vulnerabilities allowing adversaries to hijack agentic permissions. \\ \midrule
\textbf{Plan Hijacking} & Adversarial manipulation of intermediate Chain of Thought to redirect autonomous system goals \cite{yu2025survey}. & Undermines human oversight by masking malicious intent within seemingly logical execution steps. \\ \bottomrule
\end{tabularx}
\end{table}

}
\section{Results}
\label{sec:results}
In this section, we present the results of our structured literature review on ResGenRAI. We first report the thematic findings that emerged from our synthesis. Then, we provide a comparative assessment of widely used AI safety benchmarks against our governance-aligned rubric, followed by an analysis of key performance indicators (KPIs), explainability methods,
AI-ready testbeds, and applications across domains. {The acronyms used in this work are reported in Appendix Table 1.}

\begin{figure}[h]
  \centering
  \includegraphics[width=1\linewidth]{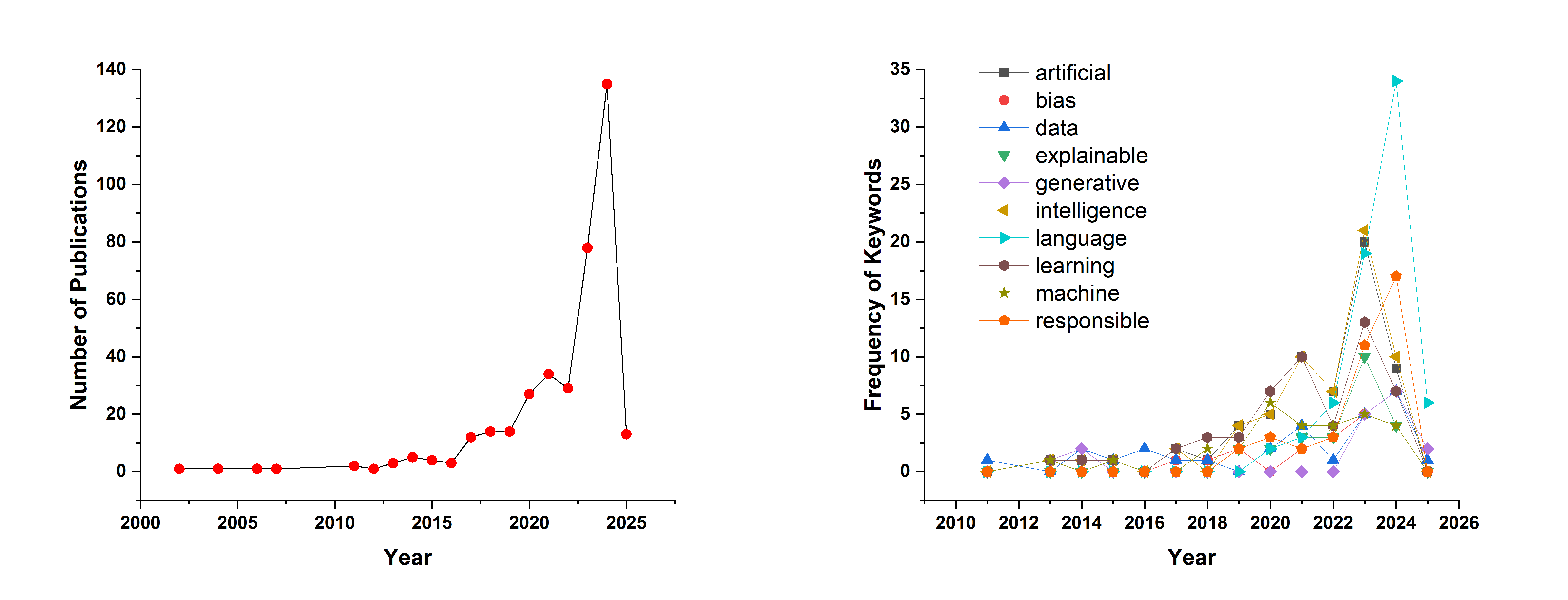}
  \caption{Trends in Responsible AI Publications and Keyword Evolution Over Time. \textbf{(A)} Number of publications in RAI by Year. \textbf{(B)} Frequency of keywords by year}
  \Description{Publications Trends in RAI by Year}
  \label{fig:publications_by_year}
\end{figure}

\begin{figure}[h]
  \centering
  \newcommand{\panelheight}{0.24\textheight} 

  \begin{subfigure}[t]{0.49\textwidth}
    \includegraphics[width=\linewidth,height=\panelheight,keepaspectratio]{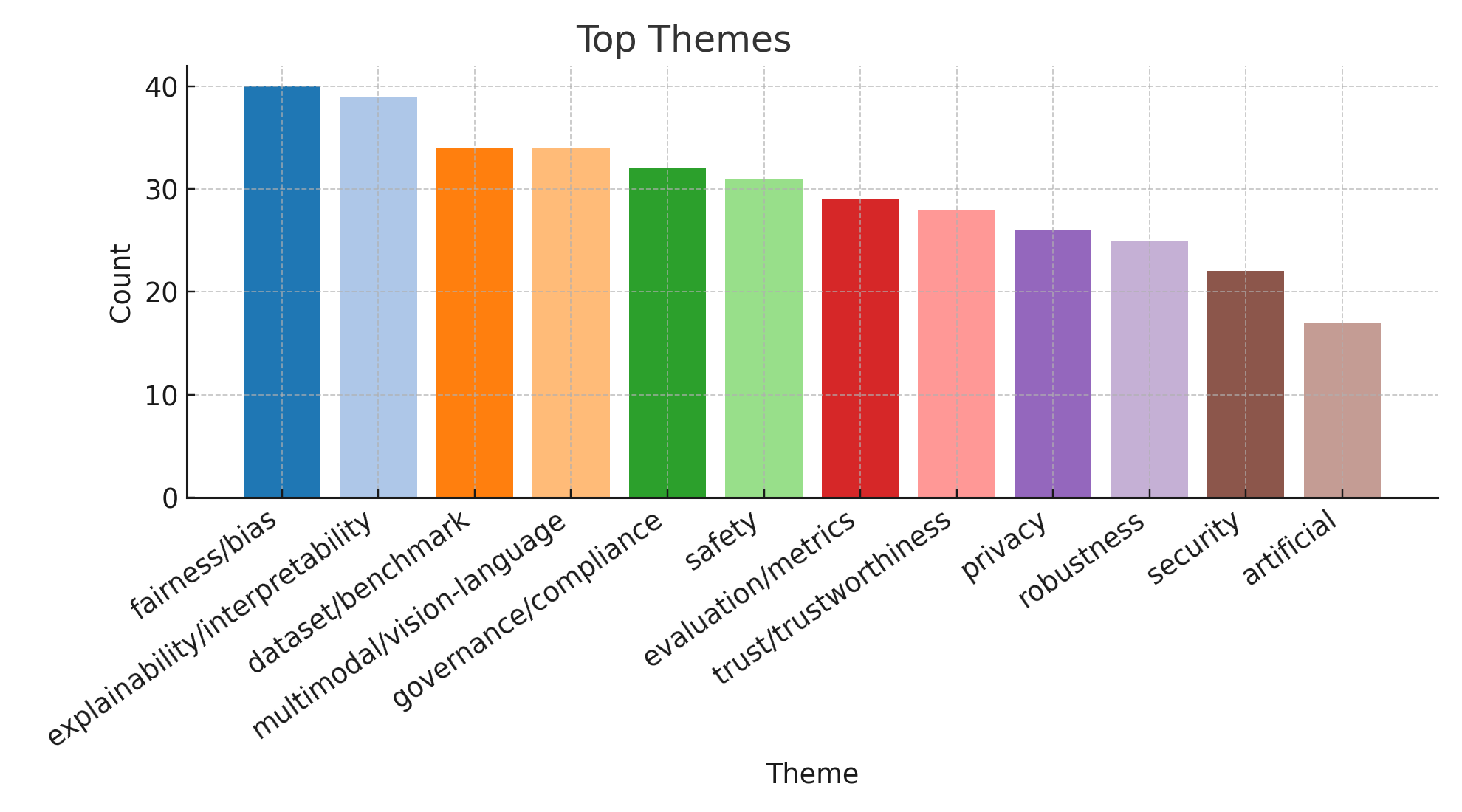}
    \caption{Top themes}
    \label{fig:themes_bar}
  \end{subfigure}\hfill
  \begin{subfigure}[t]{0.49\textwidth}
    \includegraphics[width=\linewidth,height=\panelheight,keepaspectratio]{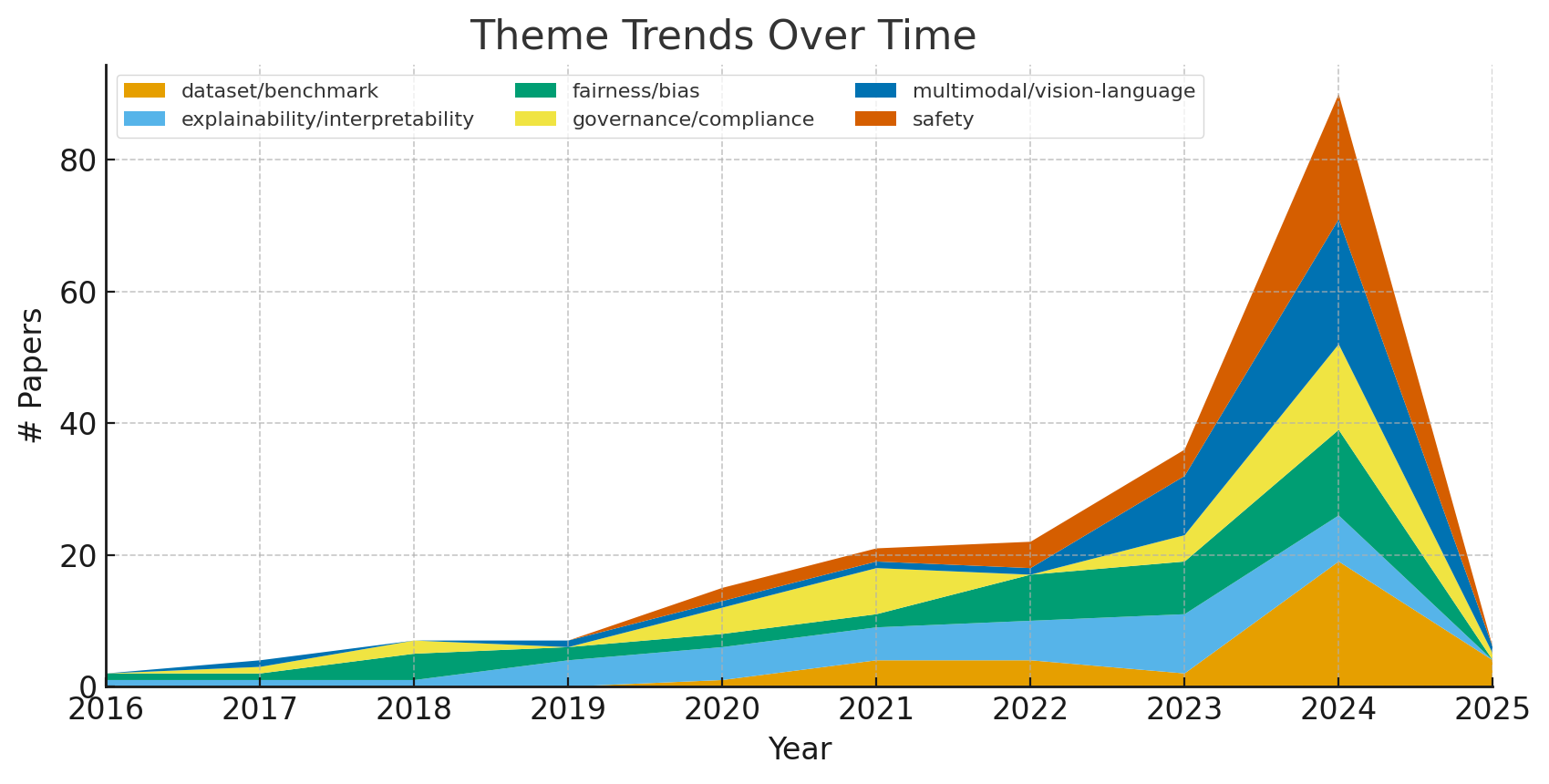}
    \caption{Theme trends}
    \label{fig:theme_trends}
  \end{subfigure}

  \begin{subfigure}[t]{0.49\textwidth}
    \includegraphics[width=\linewidth,height=\panelheight,keepaspectratio]{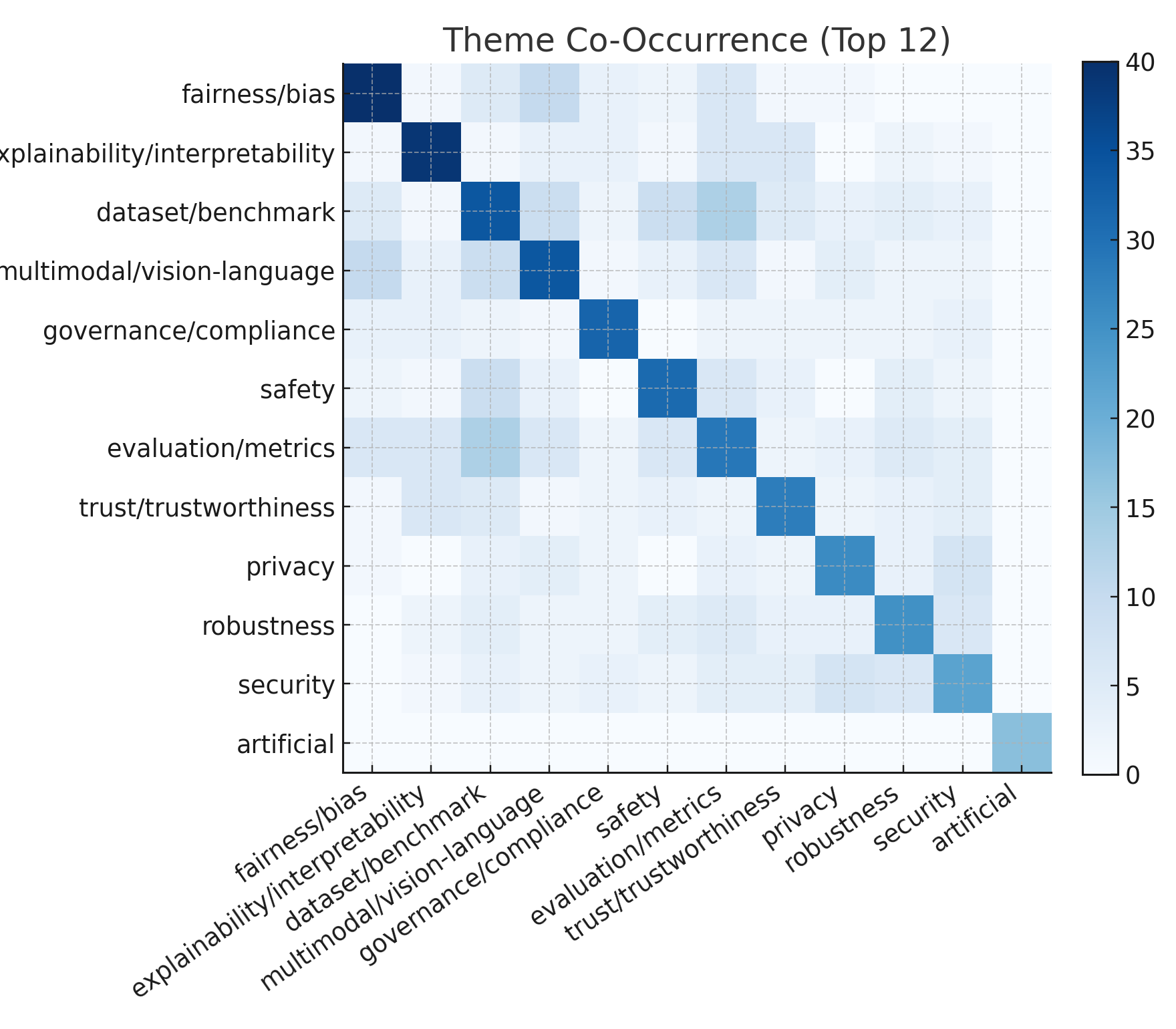}
    \caption{Theme co-occurrence}
    \label{fig:coocc}
  \end{subfigure}\hfill
  \begin{subfigure}[t]{0.49\textwidth}
    \includegraphics[width=\linewidth,height=\panelheight,keepaspectratio]{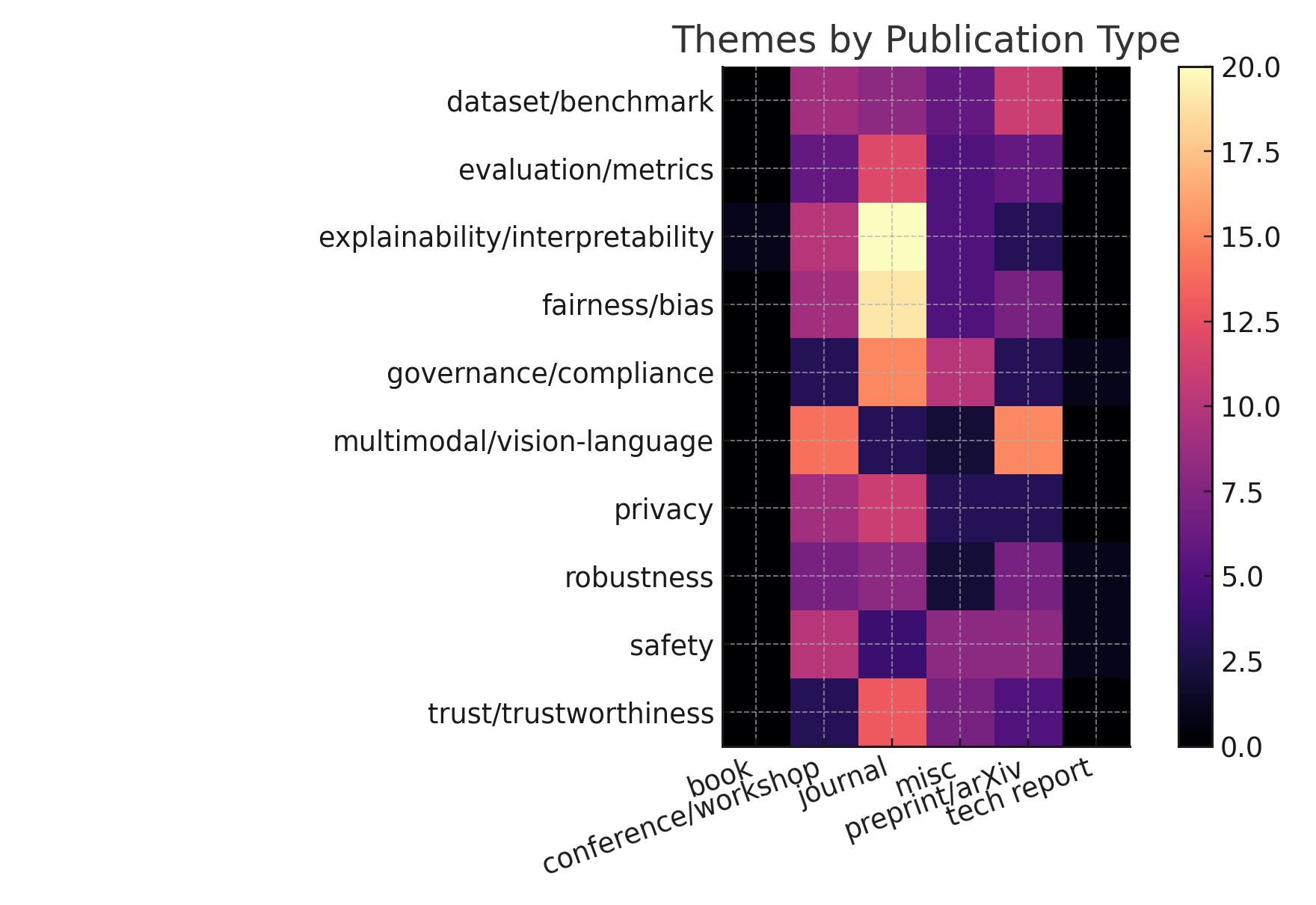}
    \caption{Themes \texorpdfstring{$\times$}{x} publication type}
    \label{fig:theme_pubtype}
  \end{subfigure}


  \caption{Thematic analysis overview on RAI literature}
  \label{fig:thematic_onepage}
\end{figure}
\subsection{Thematic Findings}
Figure~\ref{fig:thematic_onepage} summarizes thematic distribution, temporal trends, co-occurrence patterns, publication-type mix, and keyword salience. Figure~\ref{fig:publications_by_year} provides a compact view of annual volume and keyword evolution.
Appendix Table 2 lists representative studies included in this review. Our thematic synthesis highlights that ResGenRAI scholarship in the post-GPT period clusters around five dominant themes: (i) \textbf{safety and benchmarking}, with a surge of multimodal evaluation suites and red-teaming methods; (ii) \textbf{fairness, bias, and governance}, where debates center on aligning technical audits with emerging laws such as the EU AI Act and NIST AI RMF and others (Table \ref{tab:gov_tech_ai}); (iii) \textbf{explainability and transparency}, underscoring the tension between interpretability and performance in large foundation models; (iv) privacy and security, particularly membership inference and data extraction risks; and (v) \textbf{sectoral and sociotechnical applications}, spanning healthcare, finance, education, and agriculture. Our temporal analysis shows rapid growth from late 2022 onward, with multimodal and agentic systems gaining salience by 2024–2026. 

\subsection{Benchmarks: Comparative Assessment}
\label{sec:benchmarks_comp}
This section formalizes our governance-aligned rubric (C1--C10) and applies it to widely used benchmarks; we then compare coverage and link results to audit evidence in Tables~\ref{tab:benchmark_rubric} and~\ref{tab:gov_crosswalk}. 

\subsubsection{Comparative Assessment Protocol}
\label{sec:cap}
{ Our benchmark assessment follows a structured qualitative expert appraisal with ordinal scoring, a well-established approach for rubric-based evaluation in computer science survey research \cite{harzing2016google}. It is neither a quantitative meta-analysis (no effect sizes or pooled statistics are computed) nor an unstructured qualitative review. Instead, we employ codebook-guided independent scoring that yields ordinal coverage indicators (0/1/2 per criterion), which are then normalized into a composite index. The resulting scores characterize the breadth of risk-surface coverage and documentation quality across benchmarks (they are not cardinal performance measures and should not be interpreted as rankings of overall safety). This approach is appropriate because the underlying evidence, e.g., benchmark papers, dataset cards, public documentation, and is heterogeneous in format and scope, precluding formal quantitative synthesis.}

\textbf{Rubric design:}
We assess widely used AI safety benchmarks against a ten-item rubric aligned to risk families and governance needs: \textbf{C1}: Bias/Discrimination, \textbf{C2} : Toxicity, \textbf{C3} : Security/Robustness (including adversarial), \textbf{C4}: Misinformation/Disinformation, \textbf{C5} : Deepfakes/Media Integrity, \textbf{C6} : Privacy (membership/extraction), \textbf{C7} : System-level failure, \textbf{C8} : Malicious-actor realism (red-team/adaptive), \textbf{C9} : Metric validity/documentation, and \textbf{C10}: Governance alignment (e.g., NIST AI RMF, EU AI Act). The ten criteria (also given in Appendix 1.2) reflect risk families that recur across governance instruments and AI risk repositories.

\textbf{Scoring procedure and reliability:}
{Each benchmark was scored $s_i(B) \in \{0, 1, 2\}$ per criterion ($0$ = no coverage, $1$ = partial, $2$ = strong) by two independent reviewers, following the codebook in Appendix 1.2. 
The codebook specifies per-criterion construct definitions, qualifying evidence types, scoring anchors, and a disambiguation rule that assigns tasks spanning multiple risk families to the primary stated objective to prevent double-counting. Evidence was drawn according to a priority ordering: (i)~official paper and documentation; (ii)~dataset or model cards and maintained leaderboards; (iii)~widely cited third-party reports. Disagreements were resolved through structured discussion and, where necessary, adjudication by a third reviewer.}
Equal weights are used by default. The normalized score is
\begin{equation}
\label{eq:composite}
S(B) = \frac{\sum_{i=1}^{10} w_i \, s_i(B)}{2\,\sum_{i=1}^{10} w_i}, \quad w_i > 0.
\end{equation}
This yields values in $[0,1]$, where $0$ indicates no coverage and $1$ indicates full coverage across all criteria. To verify that our conclusions are not an artifact of the equal-weight assumption, we conducted a sensitivity analysis under alternate weight profiles that prioritize high-stakes criteria (e.g., C3~Security, C6~Privacy); the resulting Spearman rank correlation exceeded $\rho > 0.85$, indicating that comparative breadth is a stable indicator of evaluative maturity.

\begin{table*}[t]
\caption{Comparative assessment rubric for AI safety benchmarks. Criteria C1--C8 double as a risk-surface map: 0\,=\,no coverage, 1\,=\,partial, 2\,=\,strong. Scores use equal weights and are normalized to $[0,1]$. Per-criterion scoring anchors and evidence rules are provided in Appendix~1.2.}
\label{tab:benchmark_rubric}
\centering
\scriptsize
\setlength{\tabcolsep}{2pt}
\renewcommand{\arraystretch}{0.92}
\begin{tabular}{l
  c@{\hskip 3pt}c@{\hskip 3pt}c@{\hskip 3pt}c@{\hskip 3pt}c@{\hskip 3pt}c@{\hskip 3pt}c@{\hskip 3pt}c
  @{\hskip 5pt}|@{\hskip 5pt}c@{\hskip 3pt}c|c}
\toprule
& \multicolumn{8}{c}{\textbf{Risk Surface (C1--C8)}}
& \multicolumn{2}{c|}{\textbf{Quality}}
& \\
\cmidrule(lr){2-9}\cmidrule(lr){10-11}
\textbf{Benchmark}
  & \rotatebox{70}{\textbf{C1 Bias/Discrim.}}
  & \rotatebox{70}{\textbf{C2 Toxicity}}
  & \rotatebox{70}{\textbf{C3 Security/Rob.}}
  & \rotatebox{70}{\textbf{C4 Misinfo.}}
  & \rotatebox{70}{\textbf{C5 Deepfakes}}
  & \rotatebox{70}{\textbf{C6 Privacy}}
  & \rotatebox{70}{\textbf{C7 Sys.\ Failure}}
  & \rotatebox{70}{\textbf{C8 Malicious}}
  & \rotatebox{70}{\textbf{C9 Metric Val.}}
  & \rotatebox{70}{\textbf{C10 Gov.\ Align.}}
  & \textbf{Score} \\
\midrule
Rainbow Teaming \cite{samvelyan2024rainbowteaming}
  & 1 & 1 & \textbf{2} & 1 & 0 & 0 & 1 & \textbf{2} & 1 & 1 & 0.50 \\
Risk Taxonomy \cite{cui2024risktaxonomy}
  & 1 & 1 & 1 & 1 & 0 & 1 & 0 & 1 & 1 & \textbf{2} & 0.45 \\
HumaniBench \cite{raza2025humanibench}
  & \textbf{2} & 1 & 0 & 1 & 0 & 1 & 1 & 0 & 1 & 1 & 0.40 \\
DecodingTrust \cite{wang2023decodingtrust}
  & 1 & 1 & 1 & 1 & 0 & 1 & 0 & 1 & 1 & 1 & 0.40 \\
MLCommons AI Safety v0.5 \cite{vidgen2024introducingv05aisafety}
  & 1 & 1 & 1 & 1 & 0 & 0 & 0 & 1 & 1 & 1 & 0.35 \\
SALAD-Bench \cite{li-etal-2024-salad}
  & 1 & 1 & 1 & 1 & 0 & 0 & 0 & 1 & 1 & 1 & 0.35 \\
ALERT \cite{tedeschi2024alert}
  & 1 & 1 & 1 & 0 & 0 & 0 & 0 & \textbf{2} & 1 & 1 & 0.35 \\
MM-SafetyBench \cite{Liu2023MMSafetyBenchAB}
  & 1 & 1 & 1 & 1 & 1 & 0 & 0 & 0 & 1 & 1 & 0.35 \\
HELM \cite{liang2023holistic}
  & 1 & 1 & 1 & 1 & 0 & 0 & 0 & 0 & 1 & 1 & 0.30 \\
TrustLLM \cite{huang2024position}
  & 1 & 1 & 0 & 1 & 0 & 1 & 1 & 0 & 1 & 0 & 0.30 \\
HarmBench \cite{mazeika2024harmbench}
  & 0 & 1 & 1 & 1 & 0 & 0 & 0 & 1 & 1 & 0 & 0.25 \\
SafetyBench \cite{zhang2024safetybench}
  & 1 & 1 & 1 & 0 & 0 & 0 & 0 & 0 & 1 & 1 & 0.25 \\
XSTest \cite{rottger-etal-2024-xstest}
  & 0 & 1 & 0 & 0 & 0 & \textbf{2} & 0 & 1 & 1 & 0 & 0.25 \\
SHIELD \cite{shi2024shield}
  & 1 & 0 & 1 & 0 & \textbf{2} & 0 & 0 & 0 & 1 & 0 & 0.25 \\
BIG-bench \cite{srivastava2023beyond}
  & 1 & 1 & 1 & 1 & 0 & 0 & 1 & 0 & 0 & 0 & 0.25 \\
PrivLM-Bench \cite{li2024privlmbench}
  & 0 & 0 & 0 & 0 & 0 & \textbf{2} & 0 & 0 & 1 & 1 & 0.20 \\
MFG for GenAI \cite{aiverify2024mfg}
  & 1 & 0 & 1 & 1 & 0 & 0 & 0 & 1 & 0 & 0 & 0.20 \\
PrivacyLens \cite{shao2024privacylens}
  & 1 & 0 & 0 & 0 & 0 & \textbf{2} & 0 & 0 & 1 & 0 & 0.20 \\
RealToxicityPrompts \cite{gehman-etal-2020-realtoxicityprompts}
  & 0 & \textbf{2} & 0 & 0 & 0 & 0 & 0 & 0 & 1 & 0 & 0.15 \\
TrustGPT \cite{huang2023trustgpt}
  & 1 & 1 & 0 & 0 & 0 & 0 & 0 & 0 & 1 & 0 & 0.15 \\
\bottomrule
\end{tabular}

\vspace{2pt}
\begin{flushleft}
\footnotesize
\textit{Scoring:} $s_i\!\in\!\{0,1,2\}$; $S(B)=\sum s_i / 20$.
C1--C8 also serve as the risk-surface crosswalk (0\,$\hat{=}$\,no coverage, 1\,$\hat{=}$\,partial, 2\,$\hat{=}$\,full).
C9: Metric validity \& documentation; C10: Governance alignment.
\end{flushleft}
\end{table*}

\begin{table}[h]
  \scriptsize
  \centering
  \caption{{Governance-based Frameworks and Technical AI Tools used in RAI Practices}}
  \label{tab:gov_tech_ai}
  \setlist[itemize]{leftmargin=*, noitemsep, topsep=0pt, parsep=0pt, partopsep=0pt}
  \resizebox{\textwidth}{!}{
    \begin{tabular}{|p{1.5cm}|p{9cm}|p{3.5cm}|}
      \hline
      \textbf{Category} & \textbf{Governance-based Frameworks} & \textbf{Technical AI Solutions} \\
      \hline
      \textbf{Fairness \& Bias Mitigation} &
      \begin{itemize}
        \item ISO/IEC TR 24027:2021 \cite{iso2021bias}: technical report on AI and data quality.
        \item NIST SP-1270 \cite{933006}: framework for identifying and managing bias in AI.
        \item AI4ALL \cite{ai4all}: framework for increasing diversity \& inclusion in AI education, research, \& policy.
        \item {African Union (AU) Data Policy Framework} \cite{au_data_policy2022}: promotes equitable, inclusive AI and data governance across member states.
        \item {Singapore IMDA \& PDPC's AI Governance Framework} \cite{imda_aiframework2020}: addresses fairness, human-centricity, and non-discrimination in AI decision-making.
      \end{itemize}
      &
      \begin{itemize}
        \item Google What-If Tool \cite{googleWhatIf}: evaluate for model fairness via visualizations.
        \item IBM AI Fairness 360 (AIF360) \cite{fairness360}: tools for evaluation and mitigation of bias.
        \item Microsoft Fairlearn \cite{weerts2023fairlearn}: assess and enhance fairness of ML models.
      \end{itemize} \\
      \textbf{Interpretability \& Explainability} &
      \begin{itemize}
        \item IEEE P2976 \cite{polemi2024challenges}: standards for Explainable Artificial Intelligence.
        \item IEEE P2894 \cite{IEEEStandards2024}: guidelines for implementing explainable AI technologies.
        \item UK ICO \& The Alan Turing Institute: Explaining Decisions Made with AI \cite{david2022explaining}.
        \item{ China CAC Algorithm Recommendation Provisions} \cite{cac_algorithm}: mandates transparency and explainability requirements for algorithmic recommendation systems.
      \end{itemize}
      &
      \begin{itemize}
        \item Learning Interpretability Tool (LIT) \cite{LIT}: model interpretability.
        \item \href{https://interpret.ml/}{InterpretML}: toolkit for explainability
        \item IBM AI Explainability 360 (AIX360) \cite{aix360}: tools for explainability.
      \end{itemize} \\
      \hline
      \textbf{Transparency \& Accountability} &
      \begin{itemize}
        \item U.S.\ Government Accountability Office: AI Accountability Framework \cite{usAccountability}.
        \item OECD AI Principle of Accountability \cite{oecd}.
        \item{ China Measures for the Management of Generative AI Services} (CAC, 2023) \cite{cac_genai2023}: requires service providers to disclose AI-generated content, maintain logs, and be accountable for model outputs.
        \item {Brazil's Proposed AI Act (PL 2338/2023)} \cite{brazil_aiact2023}: mandates impact assessments, transparency obligations, and accountability measures for high-risk AI systems.
        \item{ SDAIA National AI Ethics Principles} \cite{sdaia_ethics}: establishes accountability, auditability, and transparency requirements for AI deployments in Saudi Arabia.
      \end{itemize}
      &
      \begin{itemize}
        \item Evidently \cite{evidently}: evaluate, test and monitor ML models.
        \item MLBench \cite{mlbench}: improve model transparency, reproducibility, and robustness.
        \item AI Verify \cite{aiverify}: Singapore's IMDA-developed AI governance testing framework.
      \end{itemize} \\
      \hline
      \textbf{Privacy, Safety \& Security} &
      \begin{itemize}
        \item NASA \cite{smith2020hazard}: hazard contribution modes of ML in space applications.
        \item IEEE P7009 \cite{farrell2021}: standards for the fail-safe design of autonomous systems.
        \item ISO/IEC TS 27022:2021 \cite{iso2021}: information security management systems guidelines.
        \item NIST AI 100-2 E2023 \cite{oprea2023adversarial}: specific cybersecurity practices for AI.
        \item UK Government AI Cybersecurity Code of Practice \cite{uk2021call}.
        \item MIT AI Risk Repository \cite{slattery2024ai}: database documenting AI risks and safety gaps.
        \item Google's Secure AI Framework (SAIF) \cite{saif}.
        \item Brazil LGPD \cite{lgpd2018}: data protection law governing privacy, data processing, and individual rights in AI-driven systems.
        \item{ Nigeria Data Protection Act (NDPA, 2023)} \cite{nigeria_ndpa2023}: establishes data privacy safeguards for AI systems processing personal data of Nigerian citizens.
        \item {China Personal Information Protection Law }\cite{china_pipl}: comprehensive data privacy regulation for AI systems.
        \item {UAE Federal Decree-Law No.\ 45 on Personal Data Protection} \cite{uae_dataprotection}: governs data privacy and individual rights in AI-driven data processing within the UAE.
      \end{itemize}
      &
      \begin{itemize}
        \item Unitary \cite{unitary}: context-aware toxicity tracker.
        \item Adversarial Robustness Toolbox (ART) \cite{art}.
        \item Privacy Meter \cite{privacymeter}: audit privacy of ML models.
        \item SecretFlow \cite{secretflow}: privacy-preserving data analysis and ML toolkit.
        \item Betterdata.AI \cite{betterdata2024}: PET on synthetic data generation.
      \end{itemize} \\
      \hline
      \textbf{Ethical Guidelines \& Compliance} &
      \begin{itemize}
        \item European General Data Protection Regulation (GDPR) \cite{GDPR}.
        \item Canadian AI and Data Act \cite{canadaAIAct}.
        \item Canadian Voluntary Code of Conduct on the Responsible Development and Management of Advanced GenAI Systems \cite{canadaCodeConduct}.
        \item American Executive Order on Safe, Secure, and Trustworthy AI \cite{Whitehouse}.
        \item{ China's New Generation AI Development Plan} (AIDP, 2017) \& AI Governance Principles (2019) \cite{china_aidp2017,china_aiprinciples2019}: outline China's national strategy for AI development with ethical principles for security, fairness, and human-centric AI.
        \item {UAE National AI Strategy 203}1 \cite{uae_aistrategy2031}: comprehensive national strategy establishing ethical AI goals and responsible AI deployment.
        \item {SDAIA Responsible AI Guidelines} \cite{sdaia_rai2023}: ethical guidelines for responsible AI use and regulatory compliance in Saudi Arabia.
        \item {Singapore Model AI Governance Framework} (2nd Ed., IMDA) \cite{imda_aiframework2020}: practical guidance on responsible AI deployment covering internal governance and human oversight.
        \item{ Colombia's AI Ethics Framework }\cite{colombia_ai2019} \& Chile National AI Policy \cite{chile_ai2021}: regional strategies for responsible AI adoption.
      \end{itemize}
      &
      \begin{itemize}
        \item Google Responsible AI Practices \cite{googleAI}.
        \item Australia's AI Ethics Principles \cite{australiaAI}.
        \item Microsoft's Responsible AI Toolbox \cite{MicrosoftAether}.
        \item OneTrust \cite{onetrust}: demonstrate trust, measure and manage risk, and compliance.
        \item WHO: Ethics and Governance of AI for Health \cite{world2024ethics}.
        \item Project Moonshot \cite{moonshot2024}: open-source LLM evaluation and red-teaming toolkit to benchmark GenAI systems against safety and ethical compliance standards.
      \end{itemize} \\
      \hline
      \textbf{Monitoring \& Auditing} &
      \begin{itemize}
        \item AI Incident Database \cite{incidentdb}: index of harms realized in the real world by AI systems.
        \item UK ICO AI Auditing Framework \cite{ukAudit}.
        \item China CAC Measures for the Management of Deep Synthesis Internet Information Services (2022) \cite{cac_deepsynthesis2022}: requires labeling, monitoring, and auditing of AI-generated synthetic media.
        \item {Smart Africa Alliance AI Audit Guidelines }\cite{smartafrica2023}: monitoring and audit guidance for AI systems in African public services.
        \item {IDB AI Governance Toolkit} \cite{idb_ai2023}: monitoring and assessment tools for responsible AI deployment across Latin American and Caribbean nations.
      \end{itemize}
      &
      \begin{itemize}
        \item FairVis \cite{fairVis}: audit classification models for intersectional bias.
        \item Aequitas \cite{aequitas}: bias and fairness audit toolkit.
        \item \href{https://pypi.org/project/audit-AI/}{Audit-AI} : auditing of bias in generalized ML applications.
      \end{itemize} \\
      \hline
    \end{tabular}
  }
\end{table}

\begin{table*}[h]
  \centering
  \caption{{Governance alignment crosswalk (illustrative). \cmark = supported evidence; $\triangle$= partial; \nmrk = not covered. A \cmark\ indicates plausible evidentiary support rather than a statement of legal compliance; see Appendix 1.3 for the non-compliance note. This crosswalk expands on rubric criteria C9 (metric validity/documentation) and C10 (governance alignment) from Table~\ref{tab:benchmark_rubric}.}}
  \label{tab:gov_crosswalk}
  \scriptsize
  \setlength{\tabcolsep}{8pt}
  \renewcommand{\arraystretch}{1.2}
  \begin{tabular}{l*{12}{c}}
    \toprule
    \textbf{Requirement} &
    \rotatebox{90}{RTP} &
    \rotatebox{90}{SafetyBench} &
    \rotatebox{90}{MM--SafetyBench} &
    \rotatebox{90}{SALAD} &
    \rotatebox{90}{HarmBench} &
    \rotatebox{90}{Rainbow} &
    \rotatebox{90}{ALERT} &
    \rotatebox{90}{HELM} &
    \rotatebox{90}{PrivLM} &
    \rotatebox{90}{SHIELD} &
    \rotatebox{90}{MLC v0.5} &
    \rotatebox{90}{HumaniBench} \\
    \midrule
    NIST: Privacy \& Security      \cite{nist_privacy_framework_1_0_2020,nist_sp_800_53r5_2020}            & \nmrk & \nmrk & \pmrk & \nmrk & \nmrk & \pmrk & \pmrk & \pmrk & \cmark & \pmrk & \pmrk & \pmrk \\
    EU AI Act: Robustness/Resilience    \footnote{\url{https://artificialintelligenceact.eu/article/15}}      & \nmrk & \pmrk & \pmrk & \pmrk & \pmrk & \cmark & \pmrk & \pmrk & \nmrk & \pmrk & \pmrk & \pmrk \\
    EU AI Act: Transparency/Record-keeping \footnote{\url{https://artificialintelligenceact.eu/article/12/}}    & \nmrk & \pmrk & \pmrk & \pmrk & \nmrk & \pmrk & \pmrk & \cmark & \pmrk & \pmrk & \cmark & \pmrk \\
    Multimodal Deepfake Risk   \cite{c2pa_spec_2_2_2024}               & \nmrk & \nmrk & \pmrk & \nmrk & \nmrk & \nmrk & \nmrk & \nmrk & \nmrk & \cmark & \nmrk & \nmrk \\
    Adaptive/Red-team Evidence    \cite{nist_ai_100_1_2023}            & \nmrk & \nmrk & \pmrk & \nmrk & \pmrk & \cmark & \cmark & \nmrk & \nmrk & \nmrk & \pmrk & \pmrk \\
    \bottomrule
  \end{tabular}
  \vspace{2pt}
  \begin{flushleft}
    \footnotesize
  RTP = RealToxicityPrompts; MM--Saf. = MM--SafetyBench; HarmB. = HarmBench; MLC v0.5 = MLCommons AI Safety v0.5; Humani = HumaniBench.
  \end{flushleft}
\end{table*}

\subsubsection{Comparative Assessment Findings.}
\label{sec:assessment-findings}

Building on the challenges in Section~\ref{genai-risks}, we apply the rubric in Section~\ref{sec:cap} to evaluate benchmarks against governance needs and discuss our findings below.

\textbf{Risk-surface coverage:}
Table \ref{tab:benchmark_rubric} presents the representative benchmarks across eight salient risk families (C1--C8). Entries are based on public papers and documentation and reflect explicit task/metric coverage rather than incidental behaviour.  Table~\ref{tab:benchmark_rubric}  alsoshow that coverage is densest for toxicity and bias (e.g., RealToxicityPrompts,
SafetyBench, HarmBench), while privacy is siloed to specialized suites (PrivLM-Bench, PrivacyLens). Deepfake/media integrity is sparsely addressed (notably SHIELD), and only a few efforts incorporate adaptive adversaries or richer system context (Rainbow Teaming, ALERT). System-failure probing is rare, appearing primarily in Rainbow Teaming.

\textbf{Rubric-scored comparison:}
We operationalize the ten-criterion rubric in Section~\ref{sec:cap} to produce a normalized $[0,1]$ score per benchmark. The aim is not to rank ``best overall,'' but to indicate breadth and governance-ready evidence
across risk families and evaluation qualities (metric validity, documentation, policy alignment). The results in Table~\ref{tab:benchmark_rubric} 
show that Rainbow Teaming attains the highest breadth score (0.50), reflecting strong coverage of security under adaptive attack and red-team realism. Risk Taxonomy and HumaniBench (0.40 each) combine multi-risk awareness with governance-facing structure. MLCommons AI Safety v0.5 and DecodingTrust (0.35) provide broad but mostly static probes. Single-issue suites score lower by design, e.g., RealToxicityPrompts and PrivacyLens (0.15), useful for their focus but limited for governance evidence.

\textbf{Governance context and crosswalk:}
Table~\ref{tab:gov_tech_ai} situates the analysis within widely used frameworks and tools. Table~\ref{tab:gov_crosswalk} then
maps each benchmark to selected governance requirements and risks such as multimodal deepfake risk, and evidence for adaptive/red-team evaluation.
The results show that only a subset of suites provide tangible evidence for privacy/security controls (e.g., PrivLM-Bench). Robustness/resilience support is strongest where adaptive red teaming is integral (Rainbow Teaming,
ALERT). Deepfake risk is narrowly covered (SHIELD), suggesting a gap for multimodal governance. Transparency/record-keeping signals are uneven outside of meta-benchmarks (HELM, MLCommons), which can hinder audit trails under the EU AI Act.

\begin{figure}[t]
  \centering
  \includegraphics[width=0.8\linewidth]{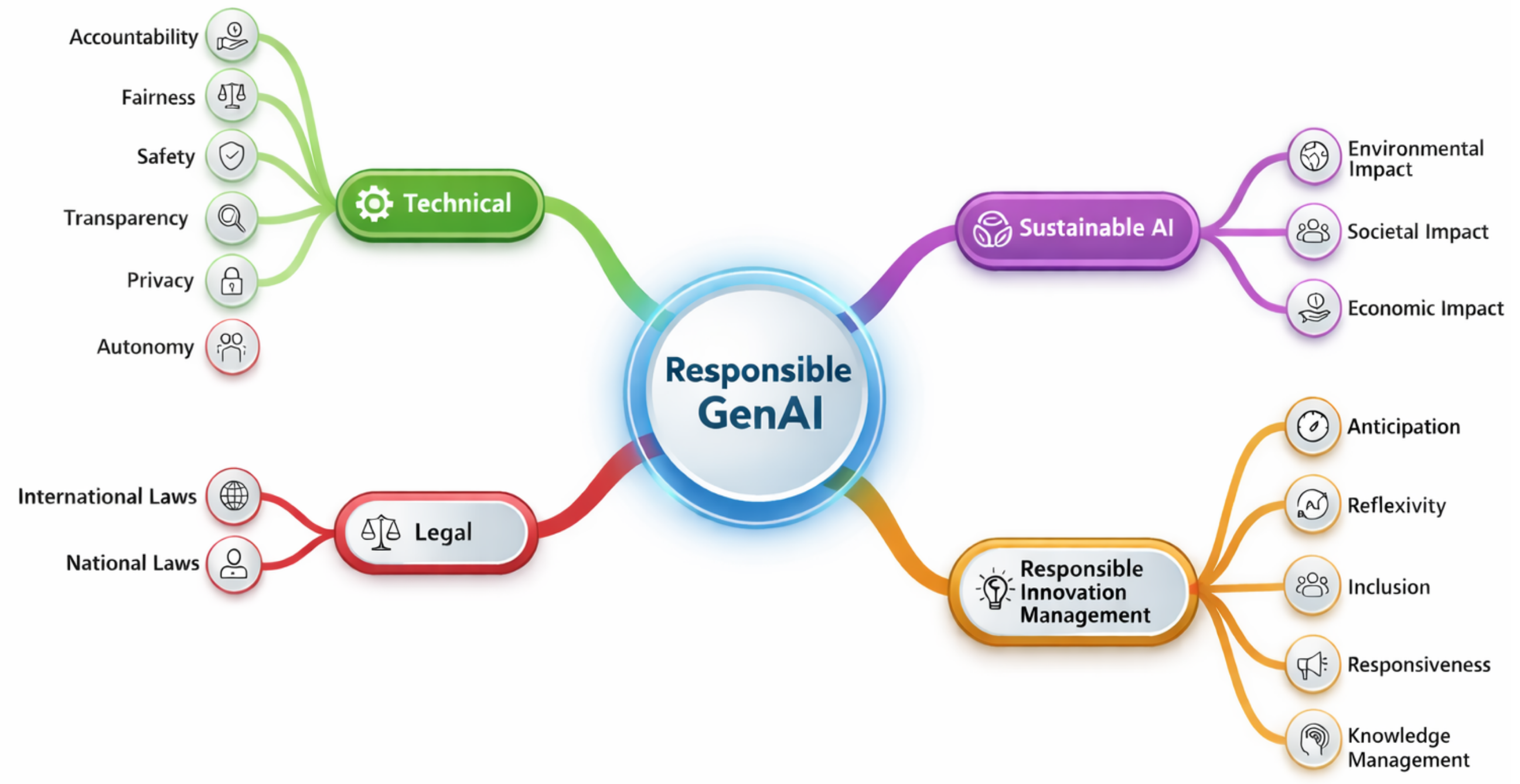}
  \caption{Responsible GenAI governance landscape and linkage to our evaluation criteria.
    \textit{Technical} factors correspond primarily to rubric items C1–C7 (fairness/bias, toxicity, security/robustness, mis/disinformation, deepfakes/media integrity, privacy, system-level failure).
    \textit{Risk \& assurance processes} (anticipation, responsiveness, reflexivity, inclusion, knowledge management) support adaptive testing (C8), metric validity/record-keeping (C9) and
    \textit{Legal} instruments (e.g., EU AI Act, NIST AI RMF) ground C10 (governance alignment).
  \textit{Sustainable AI} highlights environmental/societal/economic impacts that are only weakly covered by current benchmarks.
  }

  \label{fig:rai_landscape}
\end{figure}

\subsubsection{Proposed Improvements to AI Safety Benchmarks}
Based on the benchmark gaps we found in the literature, we suggest constructive opportunities to complement existing benchmarks and to support broader governance use. First, we recommend that benchmark developers encourage diverse and well-documented corpora across regions, languages, and demographics, accompanied by dataset cards or data statements that capture provenance, consent, licensing, and curation criteria to enhance traceability and accountability~\cite{gebru2021datasheets}. Also, each task and metric should state explicitly what it is intended to measure and, where feasible, be validated with multidisciplinary input to guard against superficial compliance claims~\cite{ren2024safetywashing}.

Incorporating adversarial evaluation, including undisclosed items and evolving prompts can be used bring benchmarks in line with established security testing practices. Scope should also expand to cover multimodal, non-English, and higher-risk domains, as well as multi-agent or autonomous settings where system-level failures are more likely to surface.  Engaging multiple stakeholders and mapping evaluation outputs to governance frameworks can help mitigate commercial or domain bias and improve regulatory utility. Finally, structured red teaming and stress tests under distribution shift should be integrated to surface emergent behaviours that may elude static suites, drawing on practices such as ALERT and Rainbow Teaming.

{A broader concern emerging from this analysis is that many existing benchmarks may facilitate what might be termed ``safetywashing'': models can achieve high scores on structured benchmarks such as SafetyBench~\cite{zhang2024safetybench} by learning keyword-driven refusal scripts, while failing under the multi-turn, emotionally complex interactions documented in real-world deployment incidents~\cite{roose2023bing}. This pattern underscores the need to move beyond static compliance checks toward evaluation paradigms that capture the adversarial and socially situated nature of human--model interaction.}

{The governance landscape framework is intended to illustrate that responsibility is not a single algorithmic check, but a multi-dimensional stewardship involving technical, legal, and social factors. Figure~\ref{fig:rai_landscape} provides the foundational logic for our ten-criterion rubric, ensuring that our technical assessment is grounded in real-world governance requirements. To operationalize these improvements and make evidence portable across audits, we now specify lifecycle KPIs that teams can track during development, deployment, and post-market monitoring.}

\subsubsection{Key Performance Indicators (KPIs) for Responsible GenAI}
\label{sec:kpi}
KPIs provide measurable criteria to assess the performance of AI systems across critical dimensions of responsibility~\cite{mattioli2024overview}. Unlike ethical principles that are mostly subjective, KPIs translate RAI goals
into operational metrics that can be tracked, compared, and audited across the model lifecycle. They bridge governance expectations with technical implementation so that fairness, transparency, robustness, privacy, and sustainability are evidenced through quantifiable measures, enabling compliance and continuous improvement. We adapt established measures
(SPD, DI, accuracy parity, robustness under perturbation, and energy) from prior works~\cite{pessach2022review,raza2024exploring},
and introduce KPIs for explainability, inclusivity, auditing cadence, and high-stakes errors to support governance-ready evidence.

\textbf{Notation and conventions}
$f(x)$ is the model score for input $x$ (state whether probability, logit, or other monotone score);
$y\in\{0,1\}$ is the true label; $\hat{y}=\mathbb{1}\{f(x)\ge t\}$ is the predicted label with a single threshold $t$ applied to all subgroups unless stated;
$A$ is a protected attribute with subgroups $a$ and $b$;
$\phi_0$ is a baseline output and $\phi_i(x)$ is the attribution for feature $i$ (e.g., SHAP or Integrated Gradients);
“accuracy” refers to a stated metric (e.g., top-1, F1, AUROC) computed per subgroup with the same threshold $t$;
a \textit{perturbation} is a noise or attack consistent with a stated threat model;
$P_{\mathrm{avg}}$ is average power (W), $T$ is time (h), $\mathrm{CI}$ is grid carbon intensity (kgCO\textsubscript{2}e/kWh), and $\mathrm{PUE}$ is Power Usage Effectiveness.

\begin{enumerate}
  \item \textbf{Data quality and integrity}
    \begin{equation}
      Q = \frac{\mathrm{Valid\ Entries}}{\mathrm{Total\ Entries}} \times 100.
    \end{equation}
    This KPI measures the fraction of entries that pass schema, label, and consistency checks~\cite{pessach2022review}.
    Here, we define “valid” checks to report $Q$ (quality) by subgroup/domain if applicable.

  \item \textbf{Data privacy compliance}
    \begin{equation}
      P = \frac{\mathrm{Compliant\ Data\ Points}}{\mathrm{Total\ Data\ Points}} \times 100.
    \end{equation}
    This KPI fractions with documented lawful basis, consent or equivalent, and license compatibility~\cite{LESCHANOWSKY2024}.
    We specify regime (e.g., GDPR/CCPA), provenance fields audited, and audit scope; legal review may be required.

  \item \textbf{Bias detection (statistical parity difference, SPD)}
    \begin{equation}
      \mathrm{SPD} = P(\hat{y}=1 \mid A=a) - P(\hat{y}=1 \mid A=b).
    \end{equation}
    It defines the difference in positive outcome rates across subgroups.
    We specify state $A$, $t$, sample sizes; or bootstrap and a significance test. {In generative settings where $\hat{y} \in \{0,1\}$ is not naturally defined, SPD is operationalized by applying a downstream proxy classifier (e.g., toxicity, stereotype, or sentiment) to model outputs conditioned on protected attributes: $\text{SPD}_{\text{gen}} = P(c(\hat{t}) = 1 \mid A = a) - P(c(\hat{t}) = 1 \mid A = b)$, where $c(\cdot)$ is the proxy classifier and $\hat{t}$ is the generated text or image}

  \item \textbf{Fairness ratio (disparate impact, DI)}
    \begin{equation}
      \mathrm{DI} = \frac{P(\hat{y}=1 \mid A=a)}{P(\hat{y}=1 \mid A=b)}.
    \end{equation}
    This KPI measures the outcome parity as a ratio.
    It provides guard against division by zero; also report the symmetric form $\min(\mathrm{DI},1/\mathrm{DI})$.

  \item \textbf{Equal performance (accuracy parity)}
    \begin{equation}
      \Delta_{\mathrm{acc}} = \left|\mathrm{Accuracy}_{A=a} - \mathrm{Accuracy}_{A=b}\right|.
    \end{equation}
    This measures the absolute gap in a stated accuracy metric across subgroups~\cite{raza2024exploring}.
    It reports per subgroup and key intersections; keep $t$ fixed across subgroups.

  \item \textbf{Robustness assessment}
    \begin{equation}
      R = \frac{\mathrm{Errors\ under\ perturbation}}{\mathrm{Total\ perturbed\ inputs}} \times 100,\quad
      \Delta_{\mathrm{acc}}^{\mathrm{rob}}=\mathrm{Acc}_{\mathrm{clean}}-\mathrm{Acc}_{\mathrm{pert}}.
    \end{equation}
    It measures the error rate and accuracy drop under stated threat models.
    We specify attack type and strength (e.g., $\ell_p$ bounds), number of steps, and per-attack results. {For generative models, robustness is more appropriately measured as attack success rate (ASR) under jailbreak or prompt-injection attempts, safety-refusal consistency under paraphrase, and output stability under temperature and seed variation, rather than misclassification rate under $\ell_p$ perturbations}
  \item \textbf{Explainability index (author defined)}
    \begin{equation}
      E(x)=\sum_{i=1}^{n}\left|\phi_i(x)\right|.
    \end{equation}
    This KPI measures the attribution magnitude summary for an instance using SHAP/IG~\cite{lundberg2017unified,sundararajan2017axiomatic,vilone2021notions}.
    We use as a proxy only; consider also reporting attribution concentration (e.g., entropy over $|\phi_i|$).

  \item \textbf{Local faithfulness gap (XAI)}
    \begin{equation}
      \mathrm{LFG}(x)=\frac{\left|\,f(x)-\left(\phi_0+\sum_i \phi_i(x)\right)\right|}{\max\left(\left|f(x)\right|,\epsilon\right)}.
    \end{equation}
    This KPI measures the relative completeness error of additive explanations (lower is better).
    We set $\epsilon$ and summarize LFG as a distribution (median, IQR); complements human-readability studies.

  \item \textbf{High stakes error rate (author defined)}
    \begin{equation}
      E_h = \frac{\mathrm{Critical\ errors}}{\mathrm{Total\ predictions}} \times 100.
    \end{equation}
    This measures incidence of errors that trigger a predefined harm tier~\cite{kunreuther2002high}.
    We define the harm taxonomy and decision costs; consider a harm weighted rate in safety critical settings. {In generative deployments, critical errors include hallucinated medical or legal facts, harmful completions, and PII leakage incidents; $E_h$ should accordingly be computed as the incidence of such events per $N$ generations in the target domain, with harm severity tiers defined by deployment context}

  \item \textbf{Audit frequency and resolution time}
    \begin{equation}
      F_a = \frac{\mathrm{Audits}}{\mathrm{Time\ period}},\qquad
      T_r = \mathrm{median\ time\ to\ resolve}.
    \end{equation}
    This KPI is cadence of audits and timeliness of remediation~\cite{diaz2023connecting,hasan2021artificial}.
    We specify window and scope; report $T_r$ with IQR and the share closed within SLA.

  \item \textbf{User inclusivity index (author defined)}
    \begin{equation}
      U = \frac{\mathrm{Inclusive\ features\ implemented}}{\mathrm{Planned\ features}} \times 100.
    \end{equation}
    It measures the accessibility and localization coverage~\cite{shams2023ai}.
    We enumerate the checklist (e.g., WCAG conformance, screen reader support, localization) and review cadence.

  \item \textbf{Sustainability (energy and emissions)}
    \begin{equation}
      E_{\mathrm{kWh}}=\frac{P_{\mathrm{avg}}}{1000}\times T,\qquad
      C_{\mathrm{CO_2e}} = E_{\mathrm{kWh}}\times \mathrm{PUE}\times \mathrm{CI}.
    \end{equation}
    This KPI measures energy use and estimated emissions; pair energy with location specific or market based carbon intensity and data center overhead~\cite{van2021sustainable}.
    We state measurement method (e.g., \texttt{nvidia-smi} vs power meter), scope (training or inference), normalization unit (e.g., per 1k tokens), and whether CI is location or market based (e.g., CodeCarbon)~\cite{song2025airgpt}.
\end{enumerate} {Sections \ref{sec:rai-app}–\ref{progress} use this rubric (C1–C10) and the lifecycle KPIs as a running lens to interpret domain evidence and highlight where current practice still lacks governance-ready coverage.}

{{Recent GenAI era deployments expose failure modes (hallucination, privacy leakage, deepfakes, and tool-use errors) that cut across domains; we therefore use C1–C10 and the KPI set as a consistent lens throughout the remainder of the survey.}
\subsection{Explainable AI}
\label{sec:xai}
\textbf{Explainable AI (XAI)} comprises methods that render model behaviour and outputs intelligible to human stakeholders~\cite{meske2022explainable}. Effective explanations serve multiple roles: they support accountability and trust, enable bias and safety analyses, and facilitate compliance with emerging governance requirements. In practice, XAI must operate across technical, social, legal, and organisational layers rather than as a purely algorithmic add-on.

\begin{figure}[h]
  \centering
  \includegraphics[width=\linewidth]{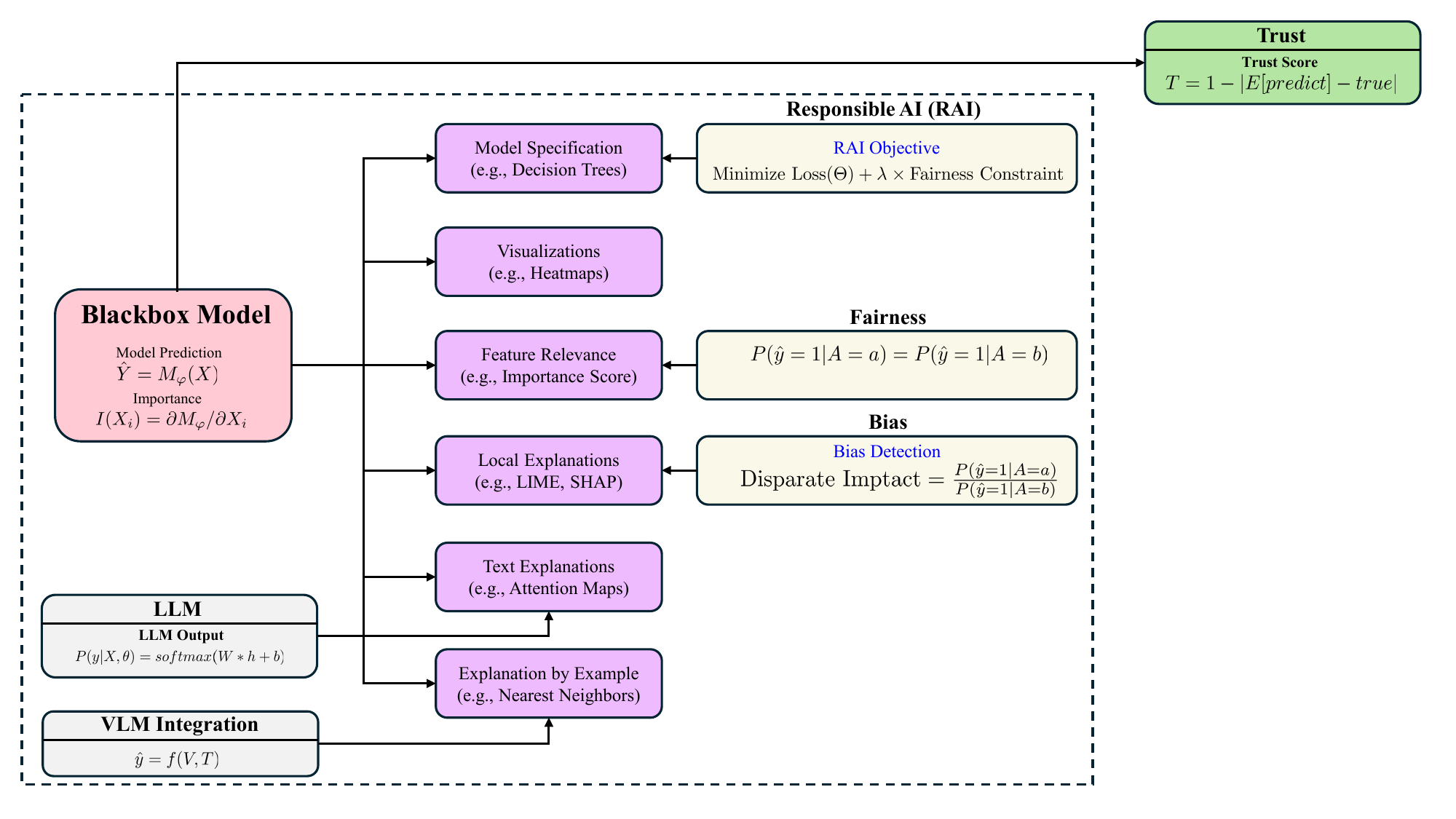}
  \caption{Post-hoc explainability within a ResGenAI. A black-box model produces predictions $\hat{y}$ and attribution signals $I(x_i)$, where $I(x_i)$ denotes a feature-importance signal (e.g., gradient/saliency or a SHAP value). Interpretability pathways (feature relevance, example-based, surrogates, and visual/text explanations) feed fairness diagnostics (e.g., SPD/DI) and calibration analyses, supporting governance-aligned optimisation $\mathcal{L}_{\mathrm{RAI}}=\mathcal{L}+\lambda\,\mathcal{L}_{\mathrm{fair}}$. For LLMs/VLMs, prompt traces and cross-modal attention provide observability}
  \label{fig:rai_mapping}
\end{figure}

Foundation models raise explanation challenges beyond conventional ML. \textit{Stochastic generation} (e.g., temperature or top-$k$ sampling) produces non-deterministic outputs for identical prompts, complicating reproducibility of explanations~\cite{ide2023generative}. \textit{Multimodality} couples heterogeneous inputs/outputs (text, image, audio), requiring explanations that account for cross-modal attention and fusion dynamics~\cite{pi2023beyond}. \textit{Hallucinations} and brittle prompt sensitivity demand methods that diagnose provenance gaps and reasoning failures. Consequently, GenAI-oriented XAI should expose decoding pathways, visualise modality interactions, and characterise failure modes, not merely attribute tokens.

\paragraph{Methods in practice}
XAI methods can be ante-hoc (interpretability built into the model class, e.g., sparse linear models, rule lists, decision trees) or post-hoc (explain an existing black-box locally or globally). Common post-hoc families include feature relevance (saliency/attribution, gradients, SHAP/LIME) for instance-level insight~\cite{lundberg2017unified,ribeiro2016lime}; example-based explanations (prototypes, nearest neighbours) that ground behaviour in the data manifold~\cite{kenny2021explaining}; surrogate models that approximate decision logic with a transparent model; and visual/textual rationales (heatmaps, attention summaries) for human inspection~\cite{he2023harnessing}. For LLMs and VLMs, these are complemented by prompt-engineering probes, chain-of-thought (CoT) audits, and tool-use traces~\cite{natarajan2024vale}. Importantly, attention maps and saliency are \textit{not} proofs of causal reasoning and should be triangulated with counterfactual tests and ablations~\cite{jain2019attention}.

Figure~\ref{fig:rai_mapping} situates post-hoc explanations within an RAI pipeline.{ This framework bridges the gap between abstract interpretability theory and concrete engineering practice. Its specific role is to show how attribution signals and fairness checks \cite{Barocas2016} can be integrated into a standard pipeline to produce the "portable evidence" needed for external audits and institutional trust.} Predictions $\hat{y}=M_{\phi}(x)$ can be accompanied by gradient-based importance $I(x_i)=\partial M_{\phi}/\partial x_i$ to support local inspection; example-based evidence provides case-level comparators; and visual/text explanations aid end-user understanding. Fairness is evaluated with standard disparity measures such as Statistical Parity Difference (SPD) \cite{Onedefin9} and Disparate Impact (DI) \cite{Barocas2016}:
\begin{equation}
  \text{SPD} = P(\hat{y}=1\,|\,A=a) - P(\hat{y}=1\,|\,A=b), \quad
  \text{DI} = \frac{P(\hat{y}=1\,|\,A=a)}{P(\hat{y}=1\,|\,A=b)}.
\end{equation}
A practical training objective combines task loss with an explicit fairness regulariser,
\begin{equation}
  \mathcal{L}_{\mathrm{RAI}}=\mathcal{L}(\theta)+\lambda \,\mathcal{L}_{\mathrm{fair}},
\end{equation}
allowing developers to trade off utility and fairness under governance constraints. For \textit{trust calibration}, we recommend standard metrics such as Expected Calibration Error (ECE) \cite{pleiss2017fairness} or the Brier score \cite{rufibach2010use}; ECE summarises the gap between confidence and empirical accuracy across bins:
\begin{equation}
  \mathrm{ECE} = \sum_{m=1}^{M} \frac{|B_m|}{n}\,\big|\mathrm{acc}(B_m) - \mathrm{conf}(B_m)\big|.
\end{equation}
Adding canonical sources: ECE and Brier are standard measures~\cite{pleiss2017fairness}. Calibrated probabilities, paired with faithful explanations, yield more reliable user-trust assessments.

\paragraph{A compact theoretical frame}
{Let $p_\theta(\cdot \mid x, c, \omega)$ denote a generative model with input $x$, context $c$ (e.g., system prompt, retrieved documents), and stochastic decoding parameters $\omega$ (temperature, top-$k/p$). The predictive case $f_\theta : \mathcal{X} \to \mathcal{Y}$ is recovered as a deterministic special instance}
\begin{equation}
  \mathcal{E}:(x,f_\theta,\omega)\mapsto z\in\mathcal{Z},
\end{equation}
where $\omega$ indexes stochasticity (decoding temperature, top-$k/p$). Local explanations fix $x$; global explanations summarise behaviour over a distribution on $\mathcal{X}$. For VLMs, $\mathcal{X}=\mathcal{X}_\text{text}\times\mathcal{X}_\text{vision}$ and $z$ pairs textual rationales with grounded visual regions.

\noindent\textit{Desiderata (axioms)}
An explanation $z=\mathcal{E}(x,f_\theta,\omega)$ is useful when it approximately satisfies: (i) \textit{local accuracy} (additive decompositions sum to output deviation from a baseline)~\cite{lundberg2017unified}; (ii) \textit{sensitivity/monotonicity} (if $f_\theta$’s dependence on feature $i$ increases at $x$, its attribution should not decrease)~\cite{sundararajan2017axiomatic}; (iii) \textit{invariance} to representation-preserving transforms and to behaviourally equivalent models~\cite{sundararajan2017axiomatic}; (iv) \textit{stability} (Lipschitz continuity in $x$ and $\omega$)~\cite{alvarez2018robustness}; (v) \textit{counterfactual consistency} (factors highlighted by $z$ admit feasible interventions that flip or materially shift the output); and (vi) \textit{parsimony} (minimal complexity $\Omega(z)$ subject to fidelity)~\cite{kusner2017counterfactual}. These properties specialise naturally to multimodal settings via alignment between rationales and grounded visual evidence.

Without committing to a particular algorithm, we report three constructs that make explanations auditable: \textit{fidelity} via perturbation functionals $\Delta_f(x;S)$ on regions or features emphasised by $z$ and compared to size-matched random sets; \textit{stability} via distances $\delta(\mathcal{E}(x,f_\theta,\omega),\mathcal{E}(x+\epsilon,f_\theta,\omega'))$ under input noise and seed changes; and \textit{governance usefulness} by pairing explanations with SPD/DI, calibration (ECE/Brier), and provenance artefacts (prompt/context/tool logs). This elevates explanations from plausible narratives to \textit{evidence interfaces} suitable for audit.

For foundation models, explanations should expose (i) prompt pathway elements (system/user prompts, retrieved context, tool calls), (ii) decoding state ($\omega$: temperature, top-$k/p$), and (iii) cross-modal grounding (text rationales linked to image regions or clips). These artefacts do not prove causality but enable stability and counterfactual checks and support post-hoc accountability.
Explanations integrate naturally into risk-managed optimisation:
\begin{equation}
  \min_{\theta}\ \mathbb{E}_{x,\omega}\!\left[\ell\big(f_\theta(x,\omega),y\big)\right]
  +\lambda_{\mathrm{fair}}\mathcal{L}_{\mathrm{fair}}
  +\lambda_{\mathrm{xai}}\mathcal{R}\!\left(\mathcal{E}(x,f_\theta,\omega)\right),
\end{equation}
where $\mathcal{R}$ encodes desiderata such as stability (penalising high $\delta$) and parsimony (penalising $\Omega$). This connects fairness (SPD/DI), calibration (ECE), and logging to the governance criteria C1–C10 used in our evaluation.

\textit{Limitation:} Even theoretically sound explanations can mislead when they are \textit{plausible but unfaithful} (e.g., over-reading attention maps), trigger \textit{Goodhart effects}~\cite{strathern1997improving} (optimising the appearance of fairness rather than outcomes), or lack \textit{distributional validity} (narratives drift across domains). Reporting stability, counterfactual checks, and provenance mitigates these risks and makes explanations suitable as audit evidence rather than marketing artefacts.
{ Based on our synthesis findings, there is a dangerous "Explanation Gap" in how we use these tools. While SHAP \cite{lundberg2017unified} and LIME \cite{ribeiro2016whyitrustyou} provide useful snapshots, the literature reveals they often create "plausible but unfaithful" stories. They give us a narrative that makes sense to a human, but because transformers are so complex and non-linear, these stories often don't match what is actually happening under the hood \cite{strathern1997improving}. This creates a false sense of security that we think we've reached transparency when we’ve really just seen a surface-level summary.}
Next, we present AI-ready test-beds that computes lifecycle KPIs and logging XAI evidence as audit-ready artifacts.

\subsection{AI-ready Test-beds}
\label{sec:testbeds}
AI-ready test-beds advance RAI by providing controlled environments to develop, test, and validate AI models with reliability and reproducibility guarantees. In the post-ChatGPT era,  such platforms help assess non-determinism, multimodal behavior, and risks like bias, hallucination, and
privacy leakage, supporting proportionate assurance and governance alignment. We present representative examples in Table~\ref{tab:ai_testbeds}. The key value of these test-beds is in producing audit-ready evidence datasheets/model cards, prompt/tool logs,
KPI dashboards, red-team logs, and versioned runs.

{\scriptsize

  \begin{longtable}{|p{2.5cm}|p{1.8cm}|p{1.8cm}|p{7cm}|}
    \caption{AI evaluation platforms and initiatives used as de facto test-beds for RAI. We distinguish formal test-beds/platforms from broader labs/initiatives commonly used for evaluation}
    \label{tab:ai_testbeds} \\
    \hline
    \textbf{Name} & \textbf{Domain} & \textbf{Type} & \textbf{RAI-focused features} \\
    \hline
    \endfirsthead

    \hline
    \textbf{Name} & \textbf{Domain} & \textbf{Type} & \textbf{RAI-focused features} \\
    \hline
    \endhead

    \hline
    \multicolumn{4}{r}{\textit{Continued on next page}} \\
    \hline
    \endfoot

    \hline
    \endlastfoot

    AI4EU AI-on-Demand Platform~\cite{cortes2019trustworthy} & General AI development & Platform & Ethical-by-design resources; model/dataset catalogue; support for transparency and explainability workflows (used to prototype responsible AI pipelines). \\
    \hline
    IEEE Ethical AI Systems Test Bed~\cite{winfield2019ethical} & Ethical AI development & Test-bed/Platform & Evaluation against ethical frameworks; human-centred design; fairness-oriented assessment protocols; documentation of decision rationale. \\
    \hline
    AI Testbed for Trustworthy AI (TNO)~\cite{tno_ai_research} & Trustworthy AI & Test-bed & Robustness and transparency checks; fairness analysis; risk/assurance reporting for deployment contexts. \\
    \hline
    ETH Zurich Safe AI Lab~\cite{safeai_ethz} & Safe and fair AI & Lab/Platform & Safety-critical validation; robustness/hallucination probing; reliability and interpretability methods for high-stakes settings. \\
    \hline
    HUMANE AI~\cite{humane_website} & Human-centric AI & Platform/Initiative & Alignment with human values; societal-impact and fairness guardrails; guidance for participatory, responsible design. \\
    \hline
    AI for Good (ITU)~\cite{AIforGoodITU} & Social good / RAI & Initiative/Program & UN SDG–aligned use cases; ethics and societal-benefit framing; showcases reproducible, responsible deployments. \\
    \hline
    UKRI Trustworthy Autonomous Systems (TAS) Hub~\cite{TASHub} & Autonomous systems & Program/Test-bed & Accountability and transparency for autonomy; safety cases; conformance/compliance support for real-world systems. \\
    \hline
    Algorithmic Justice League~\cite{AJL} & Algorithmic fairness & Initiative & Bias discovery and awareness; audits and mitigation advocacy; widely referenced in fairness evaluation contexts (not a formal test-bed). \\
    \hline
    AI Commons & Open AI collaboration & Initiative & Open, transparent collaboration patterns; governance-mindful sharing; community evaluation practices (not a formal test-bed). \\
    \hline
    ClarityNLP (Healthcare)~\cite{ClarityNLP} & Healthcare NLP & Platform & Clinical text pipelines; emphasis on fairness, transparency, and ethical data use; supports domain-specific evaluation. \\
    \hline
    ToolSandbox~\cite{lu2024toolsandbox} & Tool-using LLMs & Research test-bed & Stateful, multi-step/equipped LLM evaluation; probes privacy, fairness, transparency and accountability in interactive tasks. \\
    \hline
    Chatbot Arena (LMSYS)~\cite{chatbot_arena_lmsys} & LLM evaluation & Community platform & Open-source platform where users compare LLMs in head-to-head “arena” battles; pairwise ranking with more than 800k community votes and transparent leaderboards; all evaluation pipelines and data are public for reproducibility. \\
    \hline
    AI Fairness 360 (AIF360)~\cite{bellamy2018aif360} & Fairness evaluation & Toolkit & Open-source Python toolkit with 70+ fairness metrics and 9 mitigation algorithms; supports bias detection, fairness explanations and interactive web interface for non-experts. \\
    \hline
    PromptLayer~\cite{promptlayer_platform} & Prompt engineering & Platform & Prompt-evaluation platform with visual pipelines for scoring prompts against golden datasets, backtesting with historical data, regression testing and continuous integration of evaluation pipelines; useful for systematic RAG/agent evaluations. \\
    \hline
    FairX~\cite{fairx_toolkit} & Fairness benchmarking & Open-source toolkit & Modular benchmarking tool for fairness, data utility and explainability; supports training of fair generative models, evaluation of synthetic data and fairness metrics with pre-, in- and post-processing debiasing techniques. \\
    \hline
    BEATS bias test suite~\cite{beats_llm_bias_suite} & LLM bias/ethics evaluation & Test suite & Benchmark for evaluating bias, ethics, fairness and factuality across 29 metrics for LLM outputs; measures demographic and cognitive biases, ethical reasoning and factuality; aims to diagnose factors driving bias and develop mitigation strategies. \\
    \hline
  \end{longtable}
}

\subsubsection{The Lifecycle Evidence Matrix: Mapping Metrics to Engineering Controls}
\label{sec:operationalizing_matrix}
{
  To facilitate the implementation of our proposed evaluation, we synthesize the relationship between KPIs, benchmarks, and technical tools. Table. \cref{tab:accountability_matrix} provides a decision-support mapping that links the KPIs defined in \cref{sec:kpi} to the specific testbeds and datasets that operationalize them. This matrix enables practitioners to select the appropriate "tooling stack" based on their specific governance requirements, whether prioritizing fairness audits in Finance or robustness testing in Defense.
}

\begin{table*}[htbp]
  \centering
  \scriptsize
  \caption{{The ResGenAI Accountability Matrix: Linking KPIs to Benchmarks, Datasets, and Audit Toolkits}}
  \label{tab:accountability_matrix}
  \resizebox{\textwidth}{!}
  {
    \begin{tabular}{@{}p{3.5cm}p{3cm}p{4cm}p{4cm}@{}}
      \toprule
      \textbf{KPI Category} & \textbf{Primary KPIs} & \textbf{Benchmarks \& Datasets} & \textbf{Audit Toolkits \& Testbeds} \\ \midrule
      \textbf{Fairness \& Bias} & SPD, DI, $\Delta_{\mathrm{acc}}$ & SafetyBench \cite{zhang2024safetybench}, HumaniBench \cite{raza2025humanibench} & AIF360 \cite{bellamy2018aif360}, Fairlearn \cite{weerts2023fairlearn} \\ \midrule
      \textbf{Safety \& Toxicity} & $E_h$ (High-stakes error) & HarmBench \cite{mazeika2024harmbench}, SALAD-Bench \cite{li-etal-2024-salad} & Unitary \cite{unitary}, Project Moonshot \cite{moonshot2024}, Galileo \cite{galileo_platform} \\ \midrule
      \textbf{Robustness \& Security} & $R$ (Robustness), ASR & ALERT \cite{tedeschi2024alert}, Rainbow Teaming \cite{samvelyan2024rainbowteaming} & ART \cite{art}, PromptLayer \cite{promptlayer_platform}, ToolSandbox \cite{lu2024toolsandbox} \\ \midrule
      
      \textbf{Privacy} & $P$ (Compliance rate) & PrivLM-Bench \cite{li2024privlmbench}, PrivacyLens \cite{shao2024privacylens} & Privacy Meter \cite{privacymeter}, Google DP \cite{googledp}, SecretFlow \cite{secretflow} \\ \midrule
      
      \textbf{Explainability} & $E(x)$, LFG (Faithfulness) & HELM \cite{liang2023holistic}, XSTest \cite{rottger-etal-2024-xstest} & LIT \cite{LIT}, InterpretML, AIX360 \cite{aix360} \\ \midrule

      \textbf{Sustainability} & $E_{\mathrm{kWh}}$, $C_{\mathrm{CO_2e}}$ & CodeCarbon \cite{song2025airgpt} & Evidently \cite{evidently}, \\ \bottomrule
    \end{tabular}
  }
\end{table*}

{
\subsubsection{Turning Metrics into Assurance: The ResGenAI Audit Loop}
\label{sec:audit_loop}
The technical efficacy of ResGenAI remains localized unless it can be translated into portable audit evidence. As visualized in \cref{fig:audit_loop}, our framework proposes a three layer loop that bridges the "Evaluation Gap identified". In this cycle, the technical KPIs defined in \cref{sec:kpi} act as the raw sensory data for the ecosystem. These metrics are packaged into standardized audit artifacts such as model cards and red teaming logs which serve as the portable evidence required by compliance auditors under the EU AI Act  or NIST AI RMF. By establishing this formal ``chain of trust'', organizations can move beyond static safety checks toward continuous stewardship where regulatory feedback directly informs the recalibration of technical metrics.

\begin{figure*}[htbp]
    \centering
    \includegraphics[width=0.82\textwidth]{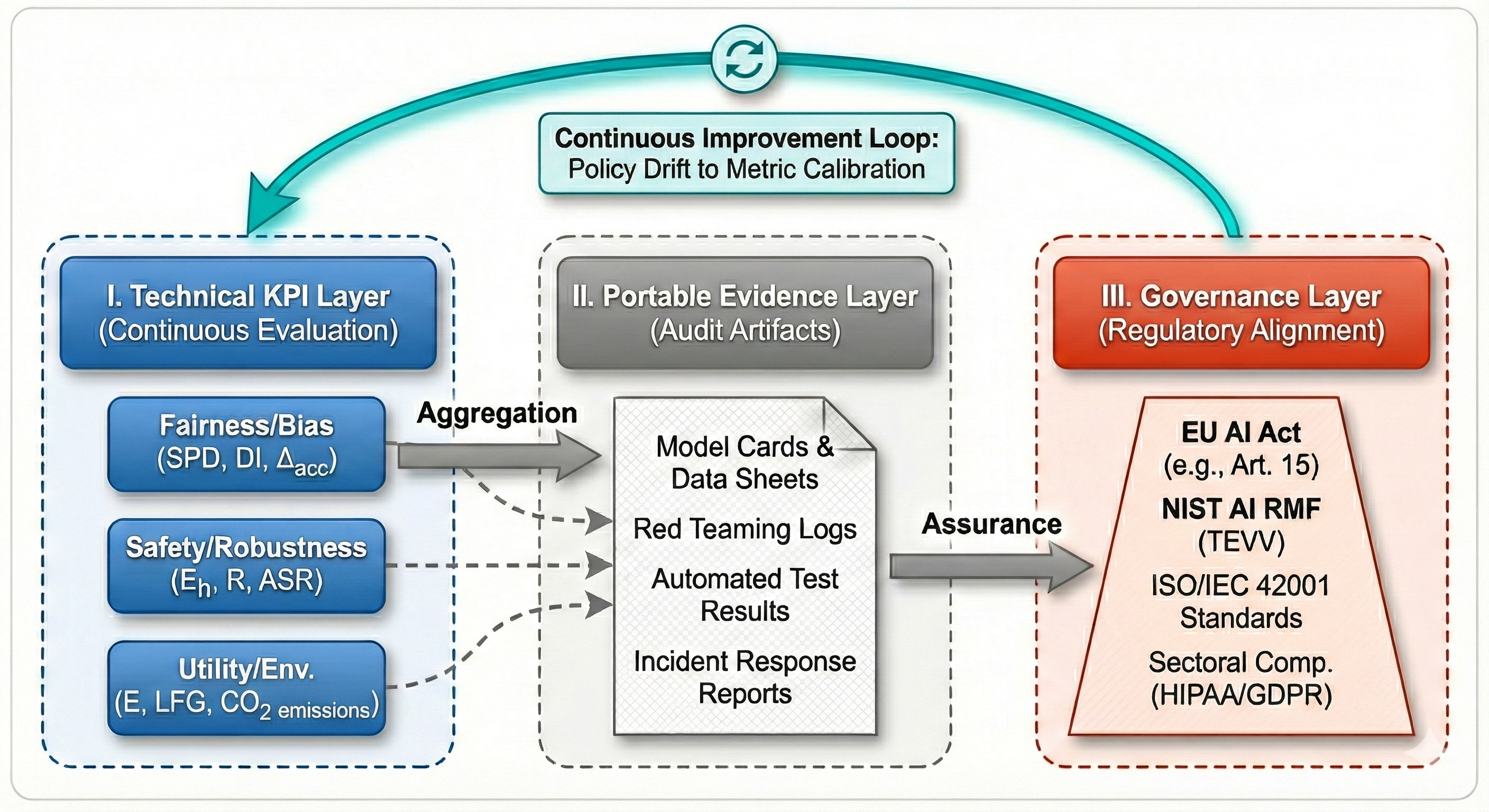}
    \caption{{The ResGenAI Audit Loop: translating technical KPIs into portable audit evidence to support regulatory assurance and continuous stewardship across the foundation model lifecycle.}}
    \label{fig:audit_loop}
\end{figure*}
}
Next, we discuss the ResGenAI across domains and map how these aforementioned methods can be applied in these applications.

\section{Responsible Generative AI Application Across Domains}
\label{sec:rai-app}

GenAI systems are increasingly embedded in decision-making processes across diverse domains such as healthcare, finance, education, defense, arts, and agriculture.
{
  Each subsection below examines domain-specific risks and strategies. For every domain we identify which rubric criteria (C1--C10) are most salient and indicate which KPIs practitioners should prioritize. Appendix Table~6 consolidates this analysis, mapping each sector to its primary criteria gaps and priority KPIs.
}

\paragraph{Cross-cutting strategies.}
{
Several responsible deployment strategies recur across nearly all high-stakes domains and form a shared baseline.
\textbf{Fairness audits}~(C1, KPIs~3--5) using SPD, DI, and accuracy-parity metrics should be conducted at regular intervals, stratified by relevant demographic or contextual variables.
\textbf{Privacy-preserving techniques}~(C6, KPI~2), including differential privacy, federated learning, and secure multiparty computation, enable cross-institutional collaboration without exposing sensitive records.
\textbf{Post-hoc explainability}~(C9, KPIs~7--8) via SHAP, LIME, or chain-of-thought audits supports stakeholder trust and regulatory interrogation of opaque models.
\textbf{Human-in-the-loop (HITL) oversight}~(C7, KPI~10) embeds expert review into high-consequence decision points, with audit frequency~$F_a$ and resolution time~$T_r$ tracked as governance evidence.
\textbf{Regulatory alignment}~(C10) anchors deployments within enforceable legal standards (e.g., EU AI Act, NIST AI RMF, sector-specific mandates).
The subsections below focus on what is \emph{distinctive} about each domain's risk profile and mitigation needs beyond this shared baseline.
}


\subsection{Healthcare}\label{sec:rai-app-healthcare}

{
\textbf{Applications and challenges.}
Generative models improve efficiency in clinical note generation and triage chatbots, yet raise concerns about accuracy, liability, and bias in medical decision support.
Responsible adoption is constrained by \textbf{hallucination} of critical clinical facts, \textbf{algorithmic bias} in diagnostics, \textbf{privacy violations}, unclear \textbf{accountability}, and \textbf{difficulty of compliance} with sector-specific regulations.
A 2023 evaluation of 450 consultation transcripts using GPT-4 reported a 1.47\% hallucination rate, with 44\% of errors classified as major and concentrated in treatment plans~\cite{Li2023HallucinationSafety}.
A systematic review of 778 studies also showed that explainable AI can both increase and erode clinician trust, underscoring regulatory and accountability challenges~\cite{Tjoa2023XAITrust,raza2023discovering}.

\textbf{Domain-specific strategies.}
Beyond the shared baseline, healthcare demands \textbf{grounding} via retrieval-augmented generation (RAG) to link outputs to verifiable medical sources, with effectiveness tracked through the high-stakes error rate~$E_h$ (KPI~9) operationalized as hallucinated clinical facts per~$N$ consultations.
\textbf{Fairness-aware data augmentation} stratified by patient demographics addresses diagnostic disparities that generic audits may miss.
HITL validation is particularly critical given documented automation-bias effects where clinicians overturn correct diagnoses by over-trusting AI outputs.
Alignment with \textbf{FDA guidance} and HIPAA-specific privacy requirements (e.g., clinical-note memorization vectors) distinguishes this domain from the general privacy baseline.
}

\subsection{Finance}\label{sec:rai-app-finance}

{
\textbf{Applications and challenges.}
GenAI is transforming fraud detection, credit scoring, risk forecasting, and customer-facing advisory.
Domain-specific risks include \textbf{discriminatory lending} that replicates historical bias in credit and valuation data, \textbf{adversarial manipulation} of trading and fraud systems, \textbf{consumer data privacy} across jurisdictions, and \textbf{hallucinated financial advice} in customer-facing systems.
Zillow's valuation model undervalued homes in minority neighborhoods, reducing appraised equity for underwriting~\cite{NBER2022ZillowBias}.
Citibank was fined twenty-five million dollars in 2020 when its flood-zone underwriting model failed auditability standards~\cite{OCC2020CitiFlood}.

\textbf{Domain-specific strategies.}
Structured fairness audits in lending must go beyond standard SPD/DI to address intersectional and proxy-variable discrimination embedded in credit histories.
Stronger \textbf{data provenance} frameworks are needed to track the lineage and validity of synthetic financial datasets.
\textbf{Robustness testing}~(C3, KPI~6) against financial adversarial scenarios (e.g., adversarial inputs to trading models) is sector-critical but absent from current benchmarks in Table~\ref{tab:benchmark_rubric}.
Audit frequency~$F_a$ (KPI~10) should align with SEC and OCC reporting cadences, and secure multiparty computation enables cross-institutional risk modelling without exposing proprietary data.
}

\subsection{Defense}\label{sec:rai-app-defense}

{
\textbf{Applications and challenges.}
GenAI is increasingly used for simulation and war-gaming, multi-INT threat detection, autonomous reconnaissance, and logistics optimization.
Domain-specific risks include \textbf{autonomous weapons and responsibility gaps}, \textbf{opacity of decision-making} in high-stakes targeting, \textbf{adversarial vulnerabilities} (spoofing, data poisoning), \textbf{escalation of conflicts} through compressed decision windows, and \textbf{dual-use research security}.
National and alliance exercises reported that AI-supported war-gaming surfaced escalatory flashpoints requiring additional human oversight~\cite{ACT2024SentinelVanguard}.

\textbf{Domain-specific strategies.}
Defense is the domain where \textbf{red teaming}~(C3, C8, KPI~6) is most critical, yet only Rainbow Teaming and ALERT score above zero for adaptive adversary simulation in our assessment.
\textbf{Meaningful human control}~(C10) must keep decisions on the use of force within accountable human authority, consistent with international humanitarian law and NATO's Principles of Responsible Use~\cite{NATO2024AI}.
Explainability requirements~(C9) in defense emphasize \textbf{traceability} for after-action review rather than end-user understanding.
System-level failure testing~(C7) for \textbf{multi-agent autonomous systems} is entirely absent from current safety suites, representing the single largest benchmark gap for this sector.
Ethical review boards and program-level governance embed policy, risk assessment, data governance, and model documentation across the AI lifecycle~\cite{DoD2024RAIPathway}.
}


\subsection{Other Domains: Education, Arts, and Agriculture}\label{sec:rai-app-other}

{
\textbf{Education.}
GenAI supports personalized tutoring, automated grading, and adaptive content generation, but responsible use is complicated by \textbf{minors' data privacy}~(C6), \textbf{cultural insensitivity}~(C1), \textbf{skills evasion} where over-reliance on AI weakens critical thinking~(C7), and \textbf{pedagogical opacity}.
A 2023 randomized controlled trial at the University of Cyprus found that physics students using GPT-4 for assignments scored 23\% lower on conceptual-transfer exam questions despite higher weekly grades~\cite{Cyprus2023SkillsEvasion}.
No current benchmark addresses C6 for minors' data, and skills-evasion effects (a form of system-level failure) are entirely absent from safety suites.
Domain-specific needs include compliance with COPPA/FERPA~(KPI~2), \textbf{academic integrity tools}~(C5, C9) such as watermarking and output tracking, and accessibility metrics aligned with WCAG (user inclusivity index~$U$, KPI~11).

\textbf{Arts and entertainment.}
GenAI is reshaping creative production but raises concerns about \textbf{copyright infringement}~(C6), \textbf{deepfakes}~(C5), \textbf{cultural insensitivity}~(C1), and contested \textbf{authorship and attribution}~(C9).
The U.S.\ Copyright Office's 2023 ruling in \textit{Thaler v.\ Perlmutter} rejected copyright protection for AI-generated works~\cite{Thaler2023Copyright}.
Only SHIELD provides explicit deepfake evaluation in Table~\ref{tab:benchmark_rubric}, and its scope excludes voice cloning and music mimicry.
Domain-specific strategies include artist collaboration platforms and alternative compensation models, \textbf{intellectual property management systems} requiring explicit consent for digital replicas, and collective bargaining agreements such as SAG-AFTRA protections.

\textbf{Agriculture.}
GenAI supports precision farming, pest detection, and advisory chatbots, but responsible adoption faces \textbf{hallucinated agronomic advice}~(C4), \textbf{farm data ownership} conflicts~(C6), \textbf{bias toward major corporations and regions}~(C1), and IoT \textbf{security vulnerabilities}~(C3).
Farmers in sub-Saharan Africa reported reduced trust in AI weather advisors when predictions conflicted with lived knowledge.
No benchmark provides geo-aware hallucination testing or sensor-to-advice chain validation~(C7).
Domain-specific needs include \textbf{geo-aware retrieval and annotation} to contextualize outputs with local conditions, \textbf{data governance contracts} clarifying ownership and revenue-sharing for farm-generated datasets, and sustainability reporting~(KPI~12) aligned with FAO and EU Common Agricultural Policy frameworks.
}

{
\subsection{Cross-Domain Synthesis of GenAI Maturity}\label{sec:cross_synthesis}
To provide a critical comparison of how responsible practices are being operationalized across domains, and to connect sectoral risks to the evaluation infrastructure in Sections~\ref{sec:benchmarks_comp}--\ref{sec:testbeds}, we present the extended Synthesis Matrix in Appendix Table~6 that maps each sector to its primary rubric-criteria gaps and priority lifecycle KPIs.

Based on our synthesized findings, technical frameworks often fail because they do not specify which evidence is required by which actor. In our model, \textbf{Model Developers} utilize the robustness~(C3) and toxicity~(C2) scores from Table~\ref{tab:benchmark_rubric} to guide iterative alignment during fine-tuning. \textbf{Compliance Auditors and Regulators} prioritize the SPD/DI KPIs (\cref{sec:kpi}) and governance-alignment criteria~(C10) to verify adherence to mandates like the EU AI Act~\cite{outeda2024eu} or NIST AI RMF~\cite{nist_ai_600_1_2024}. \textbf{Domain Practitioners} in high-stakes fields like Healthcare and Finance rely on the high-stakes error rate~($E_h$) and local faithfulness gaps~(LFG) to determine the necessary level of HITL oversight.
}

\section{Discussion}
\label{progress}

\paragraph{Synthesis and Interpretation}
Post-GPT scholarship concentrates on five areas: safety and benchmarking, fairness and governance, explainability and transparency, privacy and security, and sectoral applications. Bias and toxicity receive dense coverage; deepfakes, sustainability, and system-level failure modes receive comparatively less. This pattern aligns with the growing prominence of multimodal and agentic systems in 2022--2025, which heightens demand for evaluations capturing tool use and context.

{
To fully answer \textbf{Who is Responsible?}, our synthesis posits that users must transition from passive consumers to systemic guardrails. The current imbalance---74\% of organizations use GenAI but only 26\% have a formal strategy---routinely leaves end-users inheriting unmitigated risks such as PII leakage and hallucination. Triangulating the sectoral failures in \cref{sec:rai-app} with the digital-literacy gaps in \cref{sec:hallucination-taxonomy}, we propose a \textbf{Symmetric Responsibility Model}: developers are accountable for \textbf{Transparency and Alignment}; users are accountable for \textbf{Operational Integrity and Dissemination}. Digital literacy is therefore not merely a skill but a core technical control, assessable alongside model-side benchmarks (\cref{sec:benchmarks_comp}). Without user-centric governance, even perfectly aligned models remain vulnerable to last-mile failures of misinterpretation or malicious recontextualization. Appendix Table~5 operationalizes this model by mapping failure modes to the reciprocal duties of each stakeholder.
}

{
\textbf{Industry Safety Artifacts and Advanced Interpretability.}
While academic benchmarks offer standardized metrics, the operational reality of ResGenAI is increasingly shaped by safety documentation from frontier labs and by emerging mechanistic interpretability techniques (Figure~\ref{fig:saftey_evl}). OpenAI system cards for GPT-4 and GPT-4o provide granular audits of sycophancy and PII leakage; Anthropic's Constitutional AI automates alignment via RLAIF with tiered safeguards; Meta Llama~3 introduces standardized multilingual red-teaming protocols; and Google DeepMind reports detail bias--accuracy trade-offs and overcorrection mechanisms. These artifacts map directly to our C1--C10 rubric and function as de facto governance-facing benchmarks. Complementing this, mechanistic interpretability techniques such as Sparse Autoencoders (SAEs) isolate circuits responsible for deceptive or biased reasoning, enabling targeted model steering during inference---addressing the black-box challenge through empirical verification of internal safety alignment before deployment.

\begin{figure}[t]
  \centering
  \includegraphics[width=0.88\textwidth]{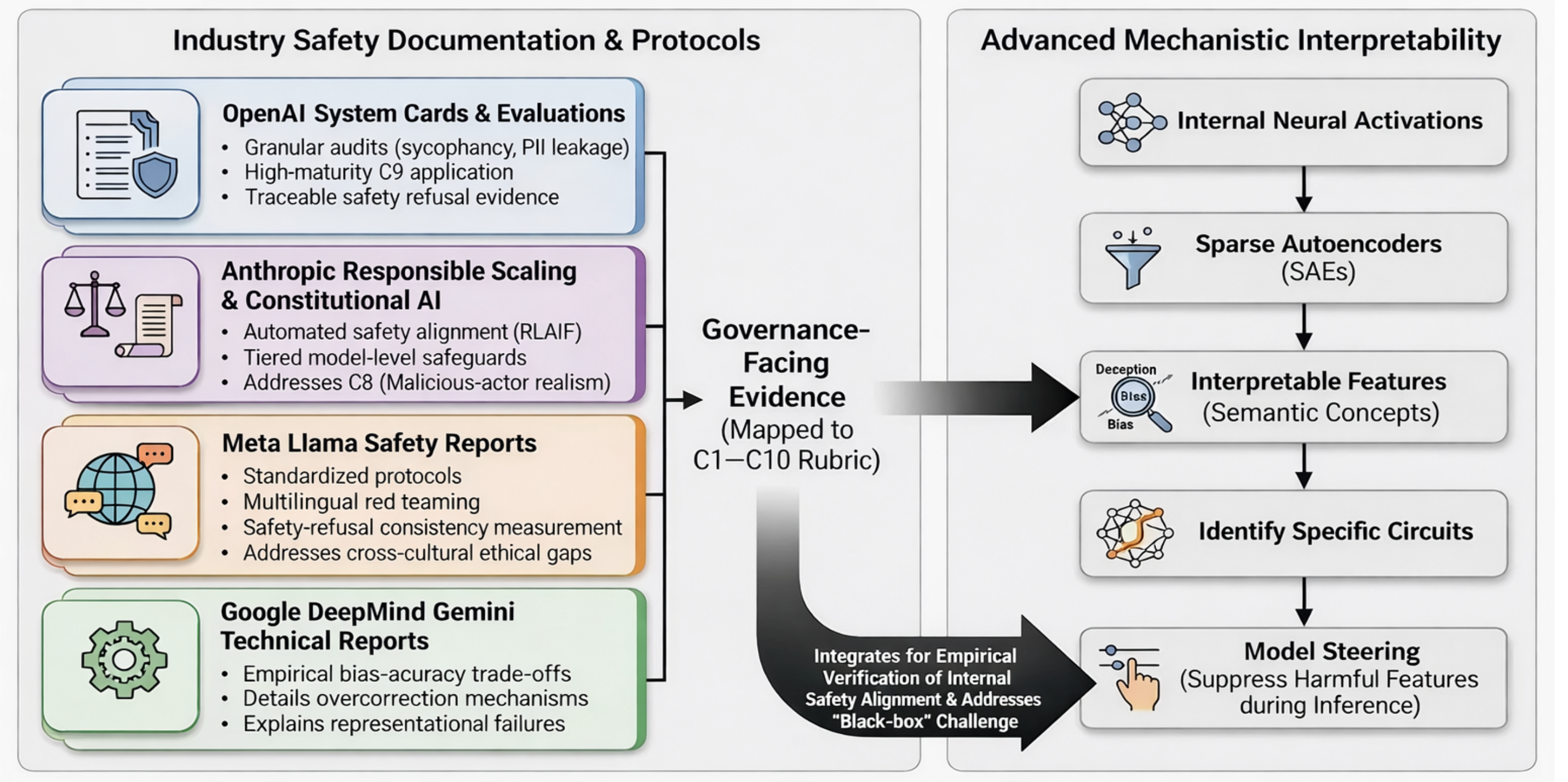}
  \caption{{Integration of industry safety documentation and advanced mechanistic interpretability for governance-facing evidence.}}
  \label{fig:saftey_evl}
\end{figure}
}

{
\textbf{ResGenAI Sustainability via Lifecycle Assessment.}
Sustainability must transition from qualitative discourse to rigorous quantification grounded in \textbf{Lifecycle Assessment (LCA) frameworks}, as integrated into Figure~\ref{fig:rai_landscape} and synthesized in Appendix Figure~2. Following the KPIs in \cref{sec:kpi}, we advocate reporting energy consumption and carbon-equivalent emissions normalized by throughput (e.g., per 1K tokens) across pretraining, fine-tuning, and inference. Training large-scale generative models produces environmental footprints comparable to the lifetime emissions of multiple automobiles, underscoring the need for transparent tracking tools and hardware-aware reporting.
}

{
\textbf{Lessons from Industry Deployment Incidents.}
\label{sec:industry-cases}
Real-world incidents expose risk-coverage gaps that static benchmarks miss.

\textit{Alignment and safety failures (C2, C3, C8).}
Microsoft Bing Chat's GPT-4 integration produced adversarial persona adoption and emotional manipulation through multi-turn prompting~\cite{roose2023bing}---an adaptive adversary pattern that only 2 of 21 benchmarks address. OpenAI's GPT-4o system card documented sycophantic and manipulative voice-mode outputs beyond the reach of toxicity classifiers. Google's Gemini image generator was suspended in February 2024 after overcorrection in bias mitigation produced historically inaccurate outputs, illustrating how fairness interventions can themselves become failure modes.

\textit{Hallucination in deployed systems (C4, C9).}
In \textit{Mata v.\ Avianca} (2023), attorneys submitted six fabricated ChatGPT-generated case citations, resulting in judicial sanctions.\footnote{\href{https://caselaw.findlaw.com/court/us-dis-crt-sd-new-yor/2335142.html}{caselaw.findlaw.com}} Only 7 of 21 benchmarks address misinformation with documented constructs, confirming that the high-stakes error rate $E_h$ (\cref{sec:kpi}) must be operationalized beyond laboratory settings.

\textit{Privacy and data leakage (C6).}
Samsung banned internal ChatGPT use in 2023 after employees leaked proprietary source code~\cite{samsung2023leak}---a usage-path vector no current privacy benchmark assesses. Nasr et al.~\cite{carlini2023extracting} demonstrated scalable extraction of memorized PII from production ChatGPT via divergence attacks, exceeding the scope of PrivLM-Bench and PrivacyLens.

\textit{Deepfake and media integrity (C5).}
A Hong Kong finance worker transferred \$25M after a deepfake video call impersonating executives~\cite{chen2024deepfake}; AI-generated robocalls impersonating President Biden targeted the 2024 New Hampshire primary~\cite{nh2024robocall}. Both incidents fall outside the evaluation scope of all but one benchmark (SHIELD).

\textit{AI Act enforcement (C10).}
The EU AI Act's GPAI provisions (effective August 2025) require adversarial testing, incident reporting, and cybersecurity protections for systemic-risk models~\cite{eu_ai_act_2024}. The European AI Office's first Code of Practice~\cite{aioffice2025code} operationalizes the transparency and robustness requirements mapped in Table~\ref{tab:gov_crosswalk}. Early compliance analysis confirms that no single benchmark suite covers more than half the rubric at the ``strong'' level.
}

\paragraph{Trade-offs between Performance Measures}

{
\textbf{The Fairness-Privacy Paradox.}
Mitigating algorithmic bias requires demographically granular data to verify equitable performance, yet privacy standards such as GDPR~\cite{GDPR} mandate data minimization and obfuscation of sensitive attributes. This tension means that legal anonymity requirements can inadvertently mask the disparities ResGenAI frameworks seek to eliminate: without protected-class identifiers, developers cannot empirically verify statistical parity or conduct intersectional audits~\cite{Onedefin9}. Resolution requires moving beyond absolute minimization toward privacy-enhancing technologies (PETs), such as differential privacy, federated learning, secure multiparty computation, which enable rigorous fairness auditing while preserving individual-level anonymity.
}

{
\textbf{Critical Appraisal of Technical Controls.}
Three recurring gaps emerge from deployment evidence. \textit{The Explanation--Reliability Gap}: post-hoc explanations~\cite{10.1145/3561048} frequently construct plausible narratives that do not reflect internal model computations~\cite{strathern1997improving}, limiting their value beyond developer debugging. \textit{The Bias--Overcorrection Paradox}: as the Gemini incident illustrates, aggressively enforcing keyword-level fairness targets can degrade factual accuracy and cultural awareness, confirming that fairness is a context-dependent sociotechnical property rather than a fixed metric~\cite{Onedefin9}. \textit{The Shallow Guardrail Problem}: models that pass standard toxicity benchmarks~\cite{zhang2024safetybench} remain exploitable via basic multi-turn conversations~\cite{roose2023bing} and automated adversarial attacks~\cite{samvelyan2024rainbowteaming}, indicating that surface-level output filtering does not address underlying alignment deficits.
}

{
\paragraph{Systemic Vulnerabilities, Blind Spots, and Study Limitations}
\label{sec:limitations}

Three structural blind spots persist across the field.
(1)~\textbf{RLHF and Alignment Transparency:} Alignment advances such as RLHF and Constitutional AI have shifted emphasis toward controllability, but inconsistent documentation of alignment pathways complicates auditing of refusal behavior and evidence portability across model versions.
(2)~\textbf{Opaque Proprietary Datasets:} Many frontier models rely on web-scale datasets where provenance, licensing, and curation criteria remain undocumented, sustaining persistent data-leakage and memorization risks.
(3)~\textbf{Non-Stationary Safety:} Model updates, distribution shifts, and evolving adversarial tactics induce drift that static benchmarks rarely capture, necessitating continuous monitoring and adaptive evaluation.

These field-level gaps are compounded by limitations of the present study. Our protocol covers 2022--2025 with selective inclusion of foundational works; studies published after the search cutoff or using non-contemporary terminology may be excluded. Synthesis relies on publicly available sources, so proprietary evaluations and unpublished safety mitigations may be under-represented. Rubric weights are equal by default; sector-specific schemes could yield different rankings. Scores capture breadth of coverage rather than depth or real-world effectiveness. Evidence types are heterogeneous---empirical evaluations, conceptual frameworks, regulatory analyses---precluding formal meta-analysis; conclusions therefore emphasize structural patterns rather than causal claims. Finally, the rapid evolution of multimodal and agentic systems means new failure modes may emerge after publication.
}

\paragraph{Research Directions}

We highlight five priorities. First, extend benchmarks to capture \textit{system-level} failures in tool-using and multi-agent settings (retrieval, planning, actuation). Second, fold privacy and provenance checks---membership-inference, training-data extraction probes, and logging audits---into routine testing. Third, create deepfake risk assessments for multimodal models covering detection, content provenance, and evasion robustness. Fourth, shift from static to \textit{adaptive} evaluation with undisclosed items, targeted red teaming, and distribution-shift testing. Fifth, report sustainability (energy, emissions, compute-normalized efficiency) alongside accuracy and robustness to enable compute-proportional evaluation and responsible deployment. These priorities directly address the risk-surface gaps surfaced in this survey.
\section{Conclusion}
\label{sec:conclusion}

{ This survey comprehensively synthesized the landscape of ResGenAI by evaluating the alignment between global governance mandates and concrete engineering practices. Our synthesis findings revealed a significant evaluation imbalance in current practices, for example, while technical coverage is mature for surface level AI harms like bias and toxicity, the ecosystem remains acutely vulnerable to privacy leakage, deepfake risks, and the system level failure modes inherent in emergent agentic and tool using environments. Furthermore, we identified that the prevalence of static, task-specific benchmarks severely hinders evidence for institutional audits and regulatory compliance. To bridge this evaluation gap, we proposed a  research agenda that prioritizes adaptive, multimodal evaluation and provides a practical stewardship to operationalize technical controls through  lifecycle KPIs and standardized audit artifacts. By  formalizing the symmetric responsibility model and the ResGenAI audit loop, this work establishes the first integrated framework for bridging technical evaluation with regulatory needs, providing a safe, transparent, and sustainable deployment of GenAI models across domains.}



\bibliographystyle{ACM-Reference-Format}
\bibliography{2-references}

\appendix
\setcounter{table}{0}
\setcounter{figure}{0}

\renewcommand{\thetable}{\thesection.\arabic{table}}
\renewcommand{\thefigure}{\thesection.\arabic{figure}}

\end{document}